\documentclass[twocolumn]{aastex63} 
\hypersetup{linkcolor=red,citecolor=green,filecolor=cyan,urlcolor=magenta}
\usepackage{multirow}
\usepackage{amsmath}
\usepackage{amssymb}

\accepted{June 25, 2021}

\shorttitle{Mid-infrared Period--Luminosity Relations for Miras}
\shortauthors{Iwanek, Soszyński \& Kozłowski}

\begin{document}

\title{Mid-Infrared Period--Luminosity Relations for Miras in the Large Magellanic Cloud}

\correspondingauthor{Patryk Iwanek}
\email{piwanek@astrouw.edu.pl}

\author[0000-0002-6212-7221]{Patryk Iwanek}
\affiliation{Astronomical Observatory, University of Warsaw, Al. Ujazdowskie 4, 00-478 Warsaw, Poland}

\author[0000-0002-7777-0842]{Igor Soszy{\'n}ski}
\affiliation{Astronomical Observatory, University of Warsaw, Al. Ujazdowskie 4, 00-478 Warsaw, Poland}

\author[0000-0003-4084-880X]{Szymon Koz{\l}owski}
\affiliation{Astronomical Observatory, University of Warsaw, Al. Ujazdowskie 4, 00-478 Warsaw, Poland}

\begin{abstract}

We present the mid-infrared (mid-IR) period--luminosity relations (PLRs) using over 1000 Mira variables in the Large Magellanic Cloud (LMC), for the four {\it Wide-Field Infrared Survey Explorer} (WISE) and the four Spitzer bands. These PLRs cover a mid-IR wavelength range from $3.4$ $\mu$m to \mbox{$22$ $\mu$m} and are presented separately for the oxygen-rich (O-rich) and carbon-rich (C-rich) Miras. These relations can be used to measure distances to individual O-rich and/or C-rich Mira stars with the accuracy of $5\%$ and $12\%$, respectively. They are the most accurate Mira PLRs in the mid-IR to date.

\end{abstract}


\section{Introduction} \label{sec:intro}

Mira variables are fundamental-mode pulsators with periods ranging from $\sim$100 days to over $1000$ days. They belong to the class of late-type, low- or intermediate-mass pulsating red giants also known as Long Period Variables (LPVs). Miras are Asymptotic Giant Branch (AGB) tracers of old- and intermediate-age populations \citep{1983ARA&A..21..271I, 2012Ap&SS.341..123W}. Like other AGB stars, they can be divided into oxygen-rich (O-rich) and carbon-rich (C-rich) giants, depending on their surface composition \citep{2010ApJ...723.1195R}. Miras, thanks to their large brightness variations, can be easily found in the Milky Way and other Local Group galaxies, and they are an extensively studied subclass of LPVs.

Since 1912, radially pulsating stars have been widely used as Galactic and extragalactic distance indicators thanks to the discovery of period--luminosity relation (PLR; \citealt{1912HarCi.173....1L}) for classical Cepheids. The history of research on PLRs for Miras dates back to the 20s of the last century and began when \citet{1928PNAS...14..963G} noticed that Miras with longer periods are on average fainter at optical wavelengths. This result has been confirmed by e.g. \citet{1942ApJ....95..248W}.

\citet{1981Natur.291..303G} were the first to report the existence of PLR for Miras at the near-infrared (NIR) wavelengths. Their research was based on 11 Miras from the Large Magellanic Cloud (LMC), and it appeared that PLR for Miras in NIR was much tighter than in the optical passbands. This NIR PLR was redefined by \citet{1989MNRAS.241..375F} and \citet{1990AJ.....99..784H}.

The availability of long-term photometric data for millions of stars from large-scale sky surveys such as \mbox{MACHO} \citep{Alcock_2000}, EROS \citep[{\it Exp\' erience de Recherche d’Objets Sombres};][]{1996VA.....40..519A}, OGLE \citep[{\it Optical Gravitational Lensing Experiment};][]{2015AcA....65....1U} allowed an in-depth study of the PLRs of LPVs. Based on the MACHO data, \citet{1999IAUS..191..151W} and then \citet{2000PASA...17...18W} showed five distinct, parallel sequences in the period--luminosity (PL) plane. It turned out that one of these sequences was composed of Miras (sequence C in Fig. 1 in \citealt{2000PASA...17...18W}).

The research on pulsating red giants conducted by the OGLE team members has significantly expanded our knowledge about PLRs for such stars. For instance, \citet{2004AcA....54..129S} showed that OGLE Small Amplitude Red Giants (\mbox{OSARGs}) form a completely different set of PL sequences than Miras and Semiregular Variables (SRVs). In turn, \citet{2005AcA....55..331S} increased the complexity of the PL distribution of LPVs and proposed a photometric method to distinguish between oxygen-rich (O-rich) and carbon-rich (C-rich) AGB stellar populations.

Many modern studies on the Mira PLRs (e.g. \citealt{2006MNRAS.369..751W, 2008MNRAS.386..313W, 2010ApJ...723.1195R, 2011MNRAS.412.2345I, 2015ApJ...807....1R, 2017EPJWC.15201009W, 2017AJ....153..170Y, 2017AJ....154..149Y, 2018AJ....156..112Y, 2019ApJ...884...20B}) showed that these stars can be used as an excellent extension of the Cosmic Distance Ladder. Due to their well-defined PLRs and high luminosity in the infrared passbands, which are less affected by interstellar extinction, Miras can be used as an attractive distance indicator (e.g. \citealt{2009MNRAS.399.1709M, 2018ApJ...865...47Q, 2018ApJ...857...67H, 2019MNRAS.482.5567M, 2020ApJ...889....5H, 2020ApJ...891...50U}).

Recently, \citet{2021arXiv210703397I} analyzed Mira variability in a wide range of wavelengths, covering $0.1$ to $40$ microns. The authors used densely-covered, accurate OGLE $I-$band light curves to create templates and fit them to NIR and mid-infrared (mid-IR) data from multiple sky surveys, extending the study to longer wavelengths (up to $40$ microns) by fitting spectral energy distributions (SEDs). As a result, they determined the variability amplitude ratios and phase lags for a range of wavelengths. Additionally, the authors presented synthetic PLRs in near- and mid-IR for the existing and future sky surveys.

The aim of this paper is to derive the mid-IR PLRs for LMC Miras discovered in the OGLE-III data \citep{2009AcA....59..239S}, using the {\it Wide Field Infrared Survey Explorer} \citep[WISE,][]{2010AJ....140.1868W} and Spitzer \citep{2004ApJS..154....1W} data, the most accurate distance to the LMC \citep{2019Natur.567..200P}, and the reddening map of the LMC \citep{2021ApJS..252...23S}.

Despite many previous works on the Mira PLRs, this subject has not been comprehensively studied in the mid-IR. Using the reddening map of the LMC and the distance to the LMC accurate to one percent, we provide the most accurate Mira PLRs to date. Additionally, for the first time, we present PLRs separately for the O-rich and C-rich Miras, in the WISE W1, W2, W3, and W4 bands.

\section{Data} \label{sec:data}

\subsection{Sample of Miras}

OGLE is one of the largest and the longest lasting time-domain variability sky surveys worldwide. Since 1992, the Galactic bulge (BLG), and later also the Magellanic Clouds (MCs) and the Galactic disk (GD) have been regularly monitored to search for variability. These observations have enabled many scientific discoveries, catalogs of variable stars, and two-band, multi-decades-long time-series photometry.

One of the products of the third phase of the OGLE project (OGLE-III, \citealt{2003AcA....53..291U}) is the catalog of LPVs in the LMC \citep{2009AcA....59..239S}. The authors discovered almost $100\;000$ LPVs, including \mbox{OSARGs}, SRVs and Mira-type stars. The two latter subclasses were separated using the $I$-band pulsation amplitude (for Miras $\Delta I$ $>$ $0.8$ mag). Using color-color ($(V-I)$ vs. $(J-K)$) and Wesenheit ($W_I$ vs. $W_{JK}$) diagrams, \citet{2009AcA....59..239S} divided Miras into O-rich and C-rich stars. This division was evaluated and confirmed by \citet{2011MNRAS.412.2345I}. In the original catalog, \citet{2009AcA....59..239S} published two-band ($I$ and $V$ in the Johnson-Cousins photometric system), 13-year-long light curves covering the time span of 1996--2009. The full catalog can be found in the on-line databases through the OGLE webpage\footnote{\url{http://ogle.astrouw.edu.pl}}

\subsection{Optical data}

In this paper, we used 1663 published LMC Miras \citep{2009AcA....59..239S} with provided coordinates, and surface chemistry classifications (1194 C-rich and 469 O-rich). The OGLE-III phase ended in 2009, and at the beginning of 2010 the OGLE project restarted its observations as the fourth phase (OGLE-IV). A detailed description of the OGLE-IV phase, instrumentation, calibration, data reduction, observations' cadence, or the sky coverage can be found in \citet{2015AcA....65....1U}.

We supplemented publicly available Mira light curves with the non-public OGLE-IV data collected over an additional 10 years since 2010. We revised their pulsation periods using two-decades-long light curves. To find periods, we used \textsc{TATRY} code based on the multiharmonic analysis of the variance algorithm \citep[ANOVA,][]{1996ApJ...460L.107S}. We visually inspected each light curve, phase-folded it with its new pulsation period, and manually removed obvious outlying points. The updated OGLE light curves have on average $\sim$1300 and $\sim$100 data points in the I-band and V-band, respectively, obtained between December 29th, 1996 and March 15th, 2020.

\subsection{Mid-IR data} \label{subsec:mid-IR}

The role of infrared observations in studying the PLRs and measuring cosmic distances is invaluable because the influence of interstellar dust on the stellar light decreases with the increasing wavelength. Moreover, Mira variables are known to undergo the mass-loss phenomenon (e.g. \citealt{2020A&A...642A..82P}), may be the cause of the formation of circumstellar dust shells around these stars. The impact of circumstellar shells on the observed stellar radiation, as for the interstellar dust, is smaller at longer wavelengths.

We cross-matched our catalog of Mira stars with two space mission databases containing mid-IR observations: WISE \citep{2010AJ....140.1868W} and Spitzer \citep{2004ApJS..154....1W}. WISE is a 40-cm diameter infrared space telescope that has monitored the entire sky in four bands $W1$ ($3.4$ $\mu$m), $W2$ ($4.6$ $\mu$m), $W3$ ($12$ $\mu$m) and $W4$ ($22$ $\mu$m). In the standard mode, the WISE telescope has observed each sky location every six months collecting several independent exposures during one flyby. The situation was slightly different in the case of the LMC area, which was observed much more frequently due to the polar trajectory of the telescope. As a result, the light curves of stars located in the LMC were covered with much denser sampling. The main mission of the WISE telescope ended in 2010, however, observations in $W1$ and $W2$ have been continued since 2011 as part of the {\it Near Earth Object WISE Reactivation Mission} \citep[NEOWISE-R][]{2011ApJ...731...53M, 2014ApJ...792...30M}. For this analysis, we retrieved the data from AllWISE Multiepoch Photometry Table\footnote{https://wise2.ipac.caltech.edu/docs/release/allwise/}, and NEOWISE-R Single Exposure (L1b) Source Table using NASA/IPAC Infrared Science Archive\footnote{https://irsa.ipac.caltech.edu/applications/Gator/}.

The Spitzer Space Telescope is an 85-cm diameter telescope with three infrared instruments onboard: {\it Infrared Array Camera} (IRAC), {\it Infrared Spectrograph} (IRS), and {\it Multiband Imaging Photometry for Spitzer} (MIPS). The LMC was observed by Spitzer in the {\it Surveying the Agents of Galaxy's Evolution} (SAGE, \citealt{2006AJ....132.2268M}) survey using IRAC in four bands: $3.6$ $\mu$m,  $4.5$ $\mu$m,  $5.8$ $\mu$m, and $8.0$ $\mu$m, also called $[3.6]$, $[4.5]$, $[5.8]$, and $[8.0]$, or $I1$, $I2$, $I3$, and $I4$, respectively. We downloaded all available measurements from the SAGE IRAC Epoch 1 and Epoch 2 Catalog using NASA/IPAC Infrared Science Archive. Light curves from these databases contain mostly two epochs.

\begin{figure}[ht!]
\centering
\includegraphics[scale=0.45]{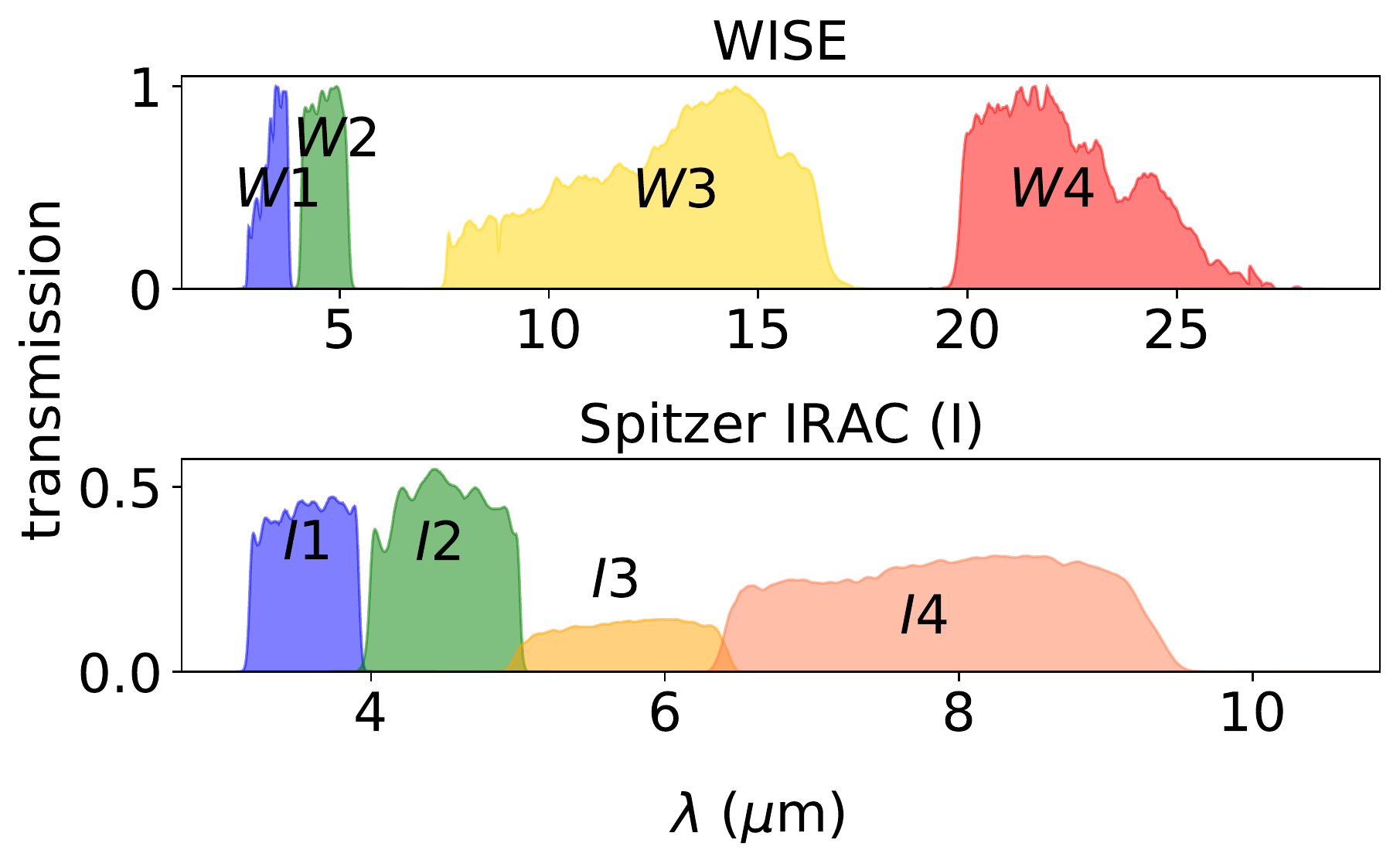}
\caption{Transmission curves for WISE and Spitzer bands.}
\label{fig:filters_transmissions}
\end{figure}

The above-mentioned databases were searched for counterparts within 1 arcsec around Miras' coordinates. 
For further analysis, we used Spitzer light curves containing two epochs in each IRAC band only. This left us with $1470$ light curves (out of 1663) collected from October to November 2005. The WISE data had to be cleared of significant outliers. After two-step cleaning procedure, we removed light curves with less than $100$ data points in the $W1$- and $W2$- 
and less than $3$ epochs in the $W3$- and $W4$-bands from further analyses. Finally, this left us with 1311 Miras observed by the WISE telescope. The median number of data points per light curve were 645, 703, 63, 22 in \mbox{$W1$-,} $W2$-, $W3$-, and $W4$-bands, respectively. The WISE data were collected between 8th February 2010 and 17th June 2010 (AllWISE data), and between 13th December 2010 and 1st December 2019 (NEOWISE-R data). A detailed description of data extraction and cleaning procedure can be found in \citet{2021arXiv210703397I}. In Figure \ref{fig:filters_transmissions}, we present transmission curves of the WISE and Spitzer bands.

The analysis of spectral energy distributions (SEDs) of the LMC Miras performed by \citet{2021arXiv210703397I} showed, that photometric measurements in the $W4-$band ($22$ $\mu$m) are typically significantly biased. While the OGLE, VMC (The VISTA survey of the Magellanic Clouds system), Spitzer, WISE ($<20$ $\mu$m) and MIPS measurements (one of the Spitzer instrument that observed the sky at $24$ $\mu$m), were well decribed by either a single (in the case of O-rich Miras) or a double (in the case of C-rich Miras) Planck function, the $W4$ measurements clearly deviated from the fit. We conducted the analysis of the PLR in the $W4-$band for completeness purposes only, but they should not be used in future analyses.

\section{Methods}
\subsection{Corrections for the interstellar extinction} \label{subsec:extinction}

Interstellar extinction has a strong influence on stellar light and should be taken into account when measuring accurate PLRs. Although this effect is smaller at longer wavelengths and the amount of dust toward the LMC is much smaller than, for example, toward the Galactic bulge, it should not be neglected.

Recently, \citet{2021ApJS..252...23S} published the most accurate optical reddening maps of the Magellanic Clouds using Red Clump (RC) stars. The reddening $E(V-I)$ and the upper $\sigma_{\mathrm{+,E(V-I)}}$ and lower $\sigma_{\mathrm{-,E(V-I)}}$ uncertainties can be obtained for a given Right Ascension (RA) and Declination (DEC) via an on-line form on the OGLE webpage\footnote{\url{http://ogle.astrouw.edu.pl/cgi-ogle/get_ms_ext.py}}. Following the authors' remarks, the extinction $A_I$ could be calculated as:

\begin{equation}
    A_\mathrm{I} \simeq 1.5 \times E(V-I),
    \label{eqn:A_I}
\end{equation}

\noindent where the coefficient can vary between 1.1 and 1.7, depending on the dust characteristics. Therefore, we used the coefficient equal to 1.5 with the uncertainty equal to 0.2, what is consistent with the coefficient $1.505$ for $A_I$ derived for $R_V = A_V/E(B-V) = 3.1$ by \citet{2011ApJ...737..103S}. Knowing that:

\begin{equation}
    E(V-I) = A_\mathrm{V} - A_\mathrm{I},
    \label{eqn:E(V-I)}
\end{equation}

\noindent the extinction in {\it V}-band is represented by the relation:

\begin{equation}
    A_\mathrm{V} = (2.5 \pm 0.2) \times E(V-I),
    \label{eqn:A_V}
\end{equation}

\noindent with the uncertainty:

\begin{equation}
    \sigma_\mathrm{A_V} = 2.5 \times E(V-I) \times \sqrt{\left(\frac{0.2}{2.5}\right)^2 + \left(\frac{\sigma_{\mathrm{+,E(V-I)}} +\sigma_{\mathrm{-,E(V-I)}} }{2\times E(V-I)}\right)^2}.
    \label{eqn:sigma_A_V}
\end{equation}

The {\it V}-band extinction could be transformed to extinctions in the WISE and Spitzer bands using extinction laws derived by \citet{2018ApJ...859..137C} and \citet{2019ApJ...877..116W}. We summarize these transformations in Table 1 for WISE W1, W2, W3, and Spitzer $[3.6]$~$\mu$m, $[4.5]$~$\mu$m, $[5.8]$ $\mu$m, $[8.0]$ $\mu$m bands. Considering the uncertainty $\sigma_{\mathrm{A_V}}$ (see Equation~(\ref{eqn:sigma_A_V})) and the transformation uncertainties $\sigma_{\mathrm{A_\lambda}/A_\mathrm{V}}$ presented in Table \ref{tab:extinction_coefficients}, the extinction uncertainties $\sigma_{\mathrm{A_\lambda}}$ in a given WISE or Spitzer band can be calculated as:

\begin{equation}
    \sigma_{\mathrm{A_\lambda}} = A_\mathrm{\lambda} \times A_\mathrm{V} \times \sqrt{\left(\frac{\sigma_{\mathrm{A_\lambda}/A_{\mathrm{V}}}}{A_{\mathrm{\lambda}}/A_{\mathrm{V}}}\right)^2 + \left(\frac{\sigma_{\mathrm{A_V}}}{A_\mathrm{V}}\right)^2}.
     \label{eqn:sigma_A_lambda}
\end{equation}

As there is a lack of information in the literature on the extinction transformation from the $V-$band to $W4-$band, we simply assumed that $A_\mathrm{W4}$ is equal $0$. \cite{2021arXiv210703397I} showed that the median $A_\mathrm{[8.0]}$ is equal to $0.007$, and is one order of magnitude smaller than extinction in $J-$band ($1.25$ $\mu$m). Knowing that the influence of interstellar dust on stellar light decreases with increasing wavelength (see e.g. \citealt{1989ApJ...345..245C}), the true extinction in $W4-$band toward the LMC must be close to $0$.

Moreover, \citet{2021arXiv210703397I} compared the extinction obtained with the method described above with the classical reddening law derived by \citet{1989ApJ...345..245C}. The authors concluded that the difference between these two methods is less than $10\%$, while extinction in the mid-IR is, in general, comparable to a typical photometric uncertainty for a survey.

The mean value of extinction uncertainties in WISE (with an exception of the $W4-$band) and Spitzer bands are presented in Table \ref{tab:extinction_calculated}. Throughout the paper, all stellar brightness and colors are extinction corrected.

\begin{table}[h!]
    \centering
    \caption{Extinction transformation coefficients between the OGLE {\it V}-band and WISE or Spitzer bands with the uncertainties. The extinction $A_{\mathrm{\lambda}}$ can be simply obtained by multiplying $A_{\mathrm{V}}$ (see Equation~(\ref{eqn:A_V})) and the coefficient $A_{\mathrm{\lambda}}/A_{\mathrm{V}}$.}
    \vspace{0.3cm}
    \begin{tabular}{l c c} 
    \hline \hline
    $\lambda$ & $A_{\mathrm{\lambda}}/A_{\mathrm{V}}$ & $\sigma_{\mathrm{A_\lambda}/A_{\mathrm{V}}}$ \\ [0.5ex] 
    \hline
    WISE W1 & 0.039 & 0.004 \\
    WISE W2 & 0.026 & 0.003 \\
    WISE W3 & 0.040 & 0.009\\
    Spitzer $[3.6]$ & 0.037 & 0.003\\
    Spitzer $[4.5]$ & 0.026 & 0.003\\
    Spitzer $[5.8]$ & 0.019 & 0.003\\
    Spitzer $[8.0]$ & 0.025 & 0.003\\
    \hline
    \label{tab:extinction_coefficients}
    \end{tabular}
\end{table}

\begin{table}[h!]
    \centering
    \caption{Mean values of calculated extinction with the uncertainties in WISE (with the exception of $W4-$band) and Spitzer bands.}
    \vspace{0.3cm}
    \begin{tabular}{l c c} 
    \hline \hline
    $\lambda$ & $A_{\lambda,\mathrm{mean}}$  & $\sigma_{\mathrm{A_{\lambda,\mathrm{mean}}}}$ \\ [0.5ex] 
    \hline
    WISE W1 & $0.012$ & $0.003$ \\
    WISE W2 & $0.008$ & $0.002$ \\
    WISE W3 & $0.012$ & $0.003$ \\
    Spitzer $[3.6]$ & $0.011$ & $0.002$ \\
    Spitzer $[4.5]$ & $0.008$ & $0.002$ \\
    Spitzer $[5.8]$ & $0.006$ & $0.001$ \\
    Spitzer $[8.0]$  & $0.008$ & $0.002$ \\
    \hline
    \label{tab:extinction_calculated}
    \end{tabular}
\end{table}

\begin{table*}[ht!]
\caption{Basic parameters for 1663 LMC Miras from the \citet{2009AcA....59..239S} catalog. We provide the coordinates, surface chemistry classification, pulsation periods, and mean magnitudes with uncertainties for all WISE and Spitzer bands. Mean magnitudes presented in this table are corrected for the interstellar extinction.}
\label{tab:miras_parameters}
\vspace{0.2cm}
\begin{center}
\begin{tabular}{lccccccccr}
\hline \hline 
Number & RA $(^h:^m:^s)$ & Decl. $(^\circ:^m:^s)$ & type & $P$ (d) & $m_{W1}$ (mag) & $\sigma_{m, W1}$ (mag) & \ldots &  $m_{[8.0]}$ (mag) & $\sigma_{m, [8.0]}$ (mag) \\ \hline \hline 
00055 & 04:29:49.87 & -70:19:00.7 & C & 289.78  & 9.315 & 0.018 & \ldots & -9.999 & -9.999 \\
00082 & 04:30:44.96 & -69:50:41.0 & O & 162.33 & -9.999 & -9.999 & \ldots & 10.959 & 0.061 \\
00094 & 04:30:54.95 & -69:28:35.5 & C & 335.87 & 10.048 & 0.017 & \ldots & 9.585 & 0.019 \\
00096 & 04:30:59.53 & -69:57:16.1 & C & 385.50 & 9.495 & 0.022 & \ldots & 7.884 & 0.035 \\
00098 & 04:31:03.28 & -69:34:15.3 & C & 323.23 & 10.064 & 0.009 & \ldots & 8.606 & 0.023 \\
00115 & 04:31:27.40 & -70:40:41.4 & C & 176.04 & -9.999 & -9.999 & \ldots & -9.999 & -9.999 \\
00144 & 04:31:54.49 & -68:42:26.3 & C & 369.67 & 9.879 & 0.014 & \ldots & 7.974 & 0.144 \\
00225 & 04:33:17.24 & -68:09:28.8 & C & 487.44 & 9.539 & 0.063 & \ldots & 6.908 & 0.061 \\
00265 & 04:33:43.68 & -70:09:50.5 & C & 444.04 & 9.885 & 0.041 & \ldots & 7.198 & 0.044 \\
\vdots & \vdots & \vdots &  \vdots & \vdots & \vdots & \vdots  & $\ddots$ & \vdots & \vdots  \\
91928 & 06:16:49.54 & -70:43:03.7 & C & 370.72 & 9.892 & 0.018 & \ldots & -9.999 & -9.999 \\
\hline \hline
\end{tabular}
\end{center}
\tablecomments{The full star ID is as in the original catalog \citep{2009AcA....59..239S} and is composed of OGLE-LMC-LPV-, folowed by the number. The table rows are sorted by the star ID. The $-9.999$ value in the columns with the mean magnitudes means that the star was not found in the mid-IR database, or the star was rejected from the analysis due to reasons described in Section \ref{sec:data}. This table is available in its entirety in a machine-readable form in the online journal. A portion is shown here for guidance regarding its form and content.}

\end{table*}

\subsection{Template light curves and mean magnitudes} \label{subsec:template_and_mean}

To measure mean magnitudes from sparsely sampled variable mid-IR light
curves, we first modeled the high-cadence
OGLE $I$-band data with the semi-parametric Gaussian process regression
(GPR) model \citep{2016AJ....152..164H}. In short, the model consists of a number of independent parts that include the mean magnitude, a low-frequency
trend across cycles, a periodic term that models the main variability,
and finally the stochastic term that absorbs any short-term
deviations. We then fitted the GPR templates by modifying their amplitude, shifting in time and magnitude to match the mid-IR light curves. Having the best parameters for the variability amplitude ratio, phase-shift, and magnitude shift,
we transformed the light curve template to the mid-IR datasets of interest.
A detailed description of making templates and fitting them to mid-IR data can be found in \citet{2021arXiv210703397I}.

We used the transformed templates to measure the mean magnitude in the WISE and Spitzer bands. Each template light curve was transformed to the flux scale, fitted with a third-order truncated Fourier series, integrated to determine the mean brightness, and finally transformed back to the magnitude scale. The uncertainties of the mean magnitude were
determined from the $\chi^2$ surface. In Table \ref{tab:miras_parameters}, we present basic parameters for each Mira star from our sample (ID, coordinates, type, pulsation period) and measured mean brightness with uncertainty in each WISE and Spitzer band.

\vspace{0.5cm}
\section{mid-IR Period--Luminosity and Period--Luminosity--Color Relations for Miras} \label{sec:PL}

PLRs for Miras are tighter and better defined at infrared wavelengths than in optical bands, as Mira PLRs are generally flat in optical (see e.g., \citealt{2021arXiv210703397I} for an explanation of this effect), but also blurred by stronger effects of the interstellar and circumstellar extinction, and hence harder to work with. The use of the Wesenheit index improves the PLRs in optical bands, however, they are still quite broad. Adding the color-term to the relations in optical bands (thus fitting period-luminosity-color relations; PLC) makes these relations tighter \citep[see e.g.][]{2019ApJ...884...20B}. In the NIR (e.g. $JHK$-bands), the PLRs for O-rich Miras are tight, while PLRs for C-rich Miras have a large scatter. For Miras of both types, PLRs are present in the mid-IR bands, but the scatter of C-rich Mira PRLs is significantly larger than that of O-rich Miras (for comparison see e.g. Figure 2 in \citealt{2011MNRAS.412.2345I} or Figure 8 in \citealt{2017AJ....154..149Y}).

In recent years, various research groups have presented different approaches to PLRs for Miras. For instance, \citet{2011MNRAS.412.2345I} compared PLRs for O-rich and C-rich Miras and concluded that the short-period Miras (with $\log{P} \lesssim 2.3$) of both types, indeed obey the same relation. On the other hand, this sample represents only a small fraction of the entire range of pulsation periods occupied by Miras.

Due to the large scatter of C-rich Mira PLRs, \citet{2011MNRAS.412.2345I} decided to analyze the O-rich Miras only. They introduced PLRs in the Spitzer mid-IR bands as a combination of two linear fits, with a kink at $\log{P} = 2.6$. \citet{2019ApJ...884...20B} found that this kink shifts to longer periods with the increasing wavelength. The authors showed that in the optical bands this kink is at $\log{P} = 2.48$, while in the NIR at $\log{P} = 2.54$. These results are consistent with \citet{2011MNRAS.412.2345I} confirming a kink at even longer periods in the mid-IR bands.

\begin{figure*}[ht!]
\centering
\includegraphics[scale=0.45]{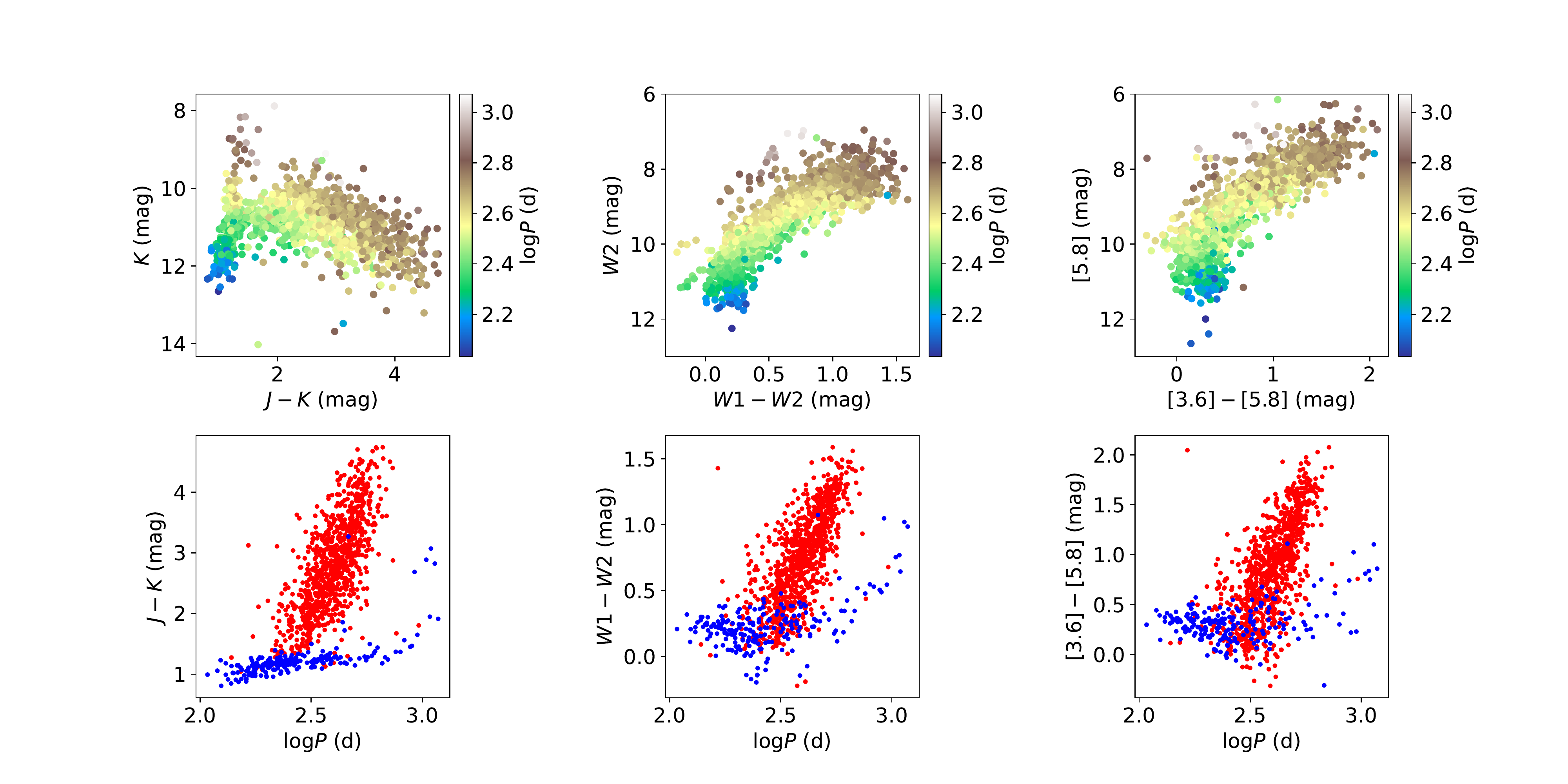}
\caption{Color-magnitude diagrams (CMDs; upper row) and period--color diagrams (lower panel) for the LMC Miras. CMDs are colored by the pulsation period, while period--color diagrams present division into C-rich (red) and O-rich (blue) Miras. Left plots present NIR color $(J-K)$, while middle and right plots present mid-IR colors, $(W1-W2)$ and $([3.6]-[5.8])$, respectively.}
\label{fig:CMDs}
\end{figure*}

Another approach to O-rich Miras assumed construction of PLRs by using a quadratic fit. For example, this approach was used by \citet{2017AJ....154..149Y} to determine PLRs for less than 180 O-rich Miras from the central region of the LMC, using time-series observations in $JHK_s$ bands. The same method was used to construct PLRs for Miras discovered in the M33 galaxy \citep{2017AJ....153..170Y, 2018AJ....156..112Y}, or to construct LMC Miras PLRs based on the Gaia DR2 data \citep{2018A&A...616A...1G, 2019ApJ...884...20B}.

Other authors proposed a linear fit to both O- and C-rich Miras separately (e.g. \citealt{2010ApJ...723.1195R, 2015ApJ...807....1R}), or even a single linear fit to both groups of Miras in the LMC (e.g. \citealt{2020ApJ...891...50U}).

\subsection{Fitting procedure} \label{subsec:models_determining}

We used the observed mean magnitudes and pulsating periods presented in Table \ref{tab:miras_parameters} to create PLRs. Our goal was to fit C-rich and O-rich Miras separately, therefore we adopted the same division as provided by \citet[][which we also present in Table \ref{tab:miras_parameters}]{2009AcA....59..239S}. The PLRs for the C-rich Miras were fitted with the linear function:

\begin{equation}
    m_{\mathrm{\lambda, fit}} = a_{\mathrm{0, \lambda}} + a_{\mathrm{1, \lambda}} \times (\log{P}-2.3),
    \label{eqn:linear_fit}
\end{equation}

\noindent where $\lambda$ is any of the WISE and Spitzer bands. 

Since the PLRs for O-rich Miras appear to be non-linear, we made both linear and quadratic fits. The quadratic fit was made for the entire range of pulsation periods and had a form:

\begin{equation}
    m_{\mathrm{\lambda, fit}} = a_{{0, \mathrm{\lambda}}} + a_{{\mathrm{1, \lambda}}} \times (\log{P}-2.3) + a_{{\mathrm{2, \lambda}}} \times (\log{P}-2.3)^2.
    \label{eqn:quadratic_fit}
\end{equation}

\noindent The O-rich Miras with pulsation periods shorter than $\sim400$ days (i.e. $\log P = 2.6$) follow the linear PLRs (e.g. \citealt{2011MNRAS.412.2345I}). Therefore, we also performed fits in the form as presented by Equation~(\ref{eqn:linear_fit}) for Miras with such pulsation periods.

We fitted models to all O-rich and C-rich Miras using the weighted least squares method, using sigma-clipping procedures, with weights that are represented by the covariance matrix $\mathcal{C}_{kl}$, and its inverse $\mathcal{B}_{kl} \equiv \mathcal{C}_{kl}^{-1}$. Following \citet{2003astro.ph.10577G} notation:

\begin{equation}
   \mathcal{C}_{kl} = \delta_{kl} \times (\sigma^2_{\mathrm{m, \lambda, k}} + \sigma_\lambda^2),
\label{eqn:matrix}
\end{equation}

\begin{equation}
   \chi^2 = \sum_{k=1}^N \sum_{l=1}^N (m_{\mathrm{\lambda}, k} -  m_{\mathrm{\lambda, fit}, k}) \mathcal{B}_{kl} (m_{\mathrm{\lambda}, l} -  m_{\mathrm{\lambda, fit}, l}), 
\label{eqn:matrix2}
\end{equation}

\noindent where $\delta_{kl}$ is the Kronecker delta ($\delta_{kl} = 1$ when $k=l$, or $\delta_{kl} = 0$, when $k \neq l$). The diagonal components of the $\mathcal{C}_{kl}$ matrix (Equation~(\ref{eqn:matrix})) contain the sum of the squares of the uncertainty of the mean magnitude $\sigma_{\mathrm{m,\lambda}}$ and the intrinsic PLR scatter $\sigma_\lambda$, which is updated iteratively during a sigma-clipping procedure. The off-diagonal elements of the matrix are equal to $0$. For each fit, we also calculated $\chi^2$ in the form as presented by Equation~(\ref{eqn:matrix2}). 

PLRs need to be calibrated using objects with known and well-measured distances. The distance to the LMC was measured with $1\%$ accuracy by \citet{2019Natur.567..200P} and it is 
$d = 49.59 \pm 0.09$ (statistical) $\pm$ 0.54 (systematic) kpc (\mbox{distance modulus $\mu = 18.477$}). We calculated the absolute zero-point $a_\mathrm{0, \lambda, abs}$ as the difference $a_\mathrm{0, \lambda} - \mu$, while the uncertainty of the absolute zero point is the sum of the zero-point $a_\mathrm{0, \lambda}$ uncertainty resulting from the fit, and both statistical and systematic uncertainties of the distance, all combined in quadrature.

\subsection{PLC Relations for C-rich Miras} \label{subsec:eqn}

A strong correlation between colors and pulsation periods of Mira 
variables exists in the optical and NIR bands. Therefore, the PLC usually has much lower scatter than the PLRs, especially for C-rich Miras \cite[see e.g.][]{2019ApJ...884...20B}. The PLC can be described in the form of the following equation:

\begin{equation}
    m_{\mathrm{\lambda, fit}} = a_{\mathrm{0, \lambda}} + a_{\mathrm{1, \lambda}} \times (\log{P}-2.3) + c \times (m_{\mathrm{\lambda_1}} - m_{\mathrm{\lambda_2}}),
    \label{eqn:plc}
\end{equation}

\noindent where $c\times(m_{\mathrm{\lambda_1}} - m_{\mathrm{\lambda_2}})$ is the color-term. In the case of optical and NIR PLC relations, the coefficient $c$ is statistically significant with relatively small uncertainty \cite[see Table 2 in][]{2019ApJ...884...20B}.

We fitted PLC relations (Equation~(\ref{eqn:plc})) to the C-rich Miras in the WISE bands using the $(W1-W2)$  color index, while in the Spitzer bands we tried fits with two color indices, i.e. $([3.6]-[4.5])$ and $([3.6]-[5.8])$. In each case, the fit including the color-term resulted in a statistically insignificant $c$ coefficient -- $c$ was close to zero with a comparable uncertainty. This led us to the conclusion that in the mid-IR the dependence between the pulsation period and color does not exist, or is very small.

To investigate this topic further, we used magnitudes in NIR bands ($J$ and $K$) extracted from the 2MASS All-Sky Catalog of Point Sources \citep{2003tmc..book.....C}. We searched for objects within 1 arcsec radius around LMC Miras' coordinates. We found 1656 (out of 1663) Miras' counterparts. We corrected $J$ and $K$ magnitudes for the interstellar extinction using the same method as described in Section \ref{subsec:extinction}, with transformations between bands derived by \citet{2018ApJ...859..137C} and \citet{2019ApJ...877..116W}.

In Figure \ref{fig:CMDs}, we present color-magnitude diagrams (CMDs, upper row) for all Miras plotted with NIR and mid-IR data mentioned above. The color map in each case presents the pulsation period of Miras. In the lower row of Figure \ref{fig:CMDs}, we present period--color diagrams using the same color-term as in CMDs. In the period--color diagrams, we plotted C-rich (red) and O-rich (blue) Miras, separately.

From the CMDs it is visible why the color-term must have an influence on the PLRs scatter in optical and NIR bands and why it is negligible in the mid-IR.  For the $K$-band, we can find very bright and blue Miras with the exact same periods as much fainter and red Miras. This is not the case for the mid-IR data, where the period seems to depend on the brightness only. 

In the left plot of the lower panel in Figure \ref{fig:CMDs}, we present period-color diagram $(J-K)$ vs. $\log P$, which is commonly used to divide AGB stars into C- and O-rich \citep{2009AcA....59..239S, 2011MNRAS.412.2345I}. It is visible that both groups of Miras form two distinct groups. A similar division is also seen for mid-IR colors (middle and right plot), but it seems that the overlap of both groups is slightly larger than that in NIR. However, the mid-IR period--color diagram could also be used for the division of AGB stars into C-rich and O-rich groups in other galaxies.

\begin{table*}[h!]
\vspace{1.8cm}
\caption{Mid-IR Period-Luminosity Relations for C-rich and O-rich Miras in WISE and Spitzer bands.}
\vspace{0.3cm}
\begin{center}
\begin{tabular}{lccccccccr}
\hline \hline
\multicolumn{10}{c}{C-rich Miras$^a$}      \\
\hline \hline
$\lambda$            & $a_{\mathrm{0, \lambda}}$  & $a_{\mathrm{0, \lambda, abs}}^{\dagger}$                             & $a_{\mathrm{1, \lambda}}$                               & $a_{\mathrm{2, \lambda}}$                            & $\sigma_{\mathrm{\lambda}}^c$ & $\chi^2/\mathrm{dof}^d$ &  $N_{\mathrm{in}}$                                       & $N_{\mathrm{fin}}$ & $N_{\mathrm{out}}$ \\ \hline
WISE W1 (Fig. \ref{fig:pl_wise_c_rich}a)      & $11.006 \pm 0.023$ & $-7.471 \pm 0.033$ & $-4.198 \pm 0.074$  & $0.000$                              & $0.258$ & $0.97$           & $999$ & $982$ & $17$  \\
WISE W2  (Fig. \ref{fig:pl_wise_c_rich}b)         & $10.916 \pm 0.030$ & $-7.561 \pm 0.038$ & $-6.445 \pm 0.094$  & $0.000$                              & $0.326$ & $0.97 $               & $998$                      & $986$ & $12$   \\
WISE W3  (Fig. \ref{fig:pl_wise_c_rich}c)         & $10.545 \pm 0.052$ & $-7.932 \pm 0.058$ & $-9.040 \pm 0.164$ & $0.000$                              & $0.584$ & $0.99$               & $1018$                     & $1012$ & $6$    \\ \hline
Spitzer $[3.6]$ (Fig. \ref{fig:pl_spitzer_c_rich}a)& $10.734 \pm 0.028$ & $-7.743 \pm 0.037$ & $-4.147 \pm 0.091$  & $0.000$                              & $0.344$ & $0.89 $               & $1072$                     & $1064$ & $8 $  \\
Spitzer $[4.5]$ (Fig. \ref{fig:pl_spitzer_c_rich}b)& $10.775 \pm 0.033$ & $-7.702 \pm 0.041$ & $-5.780 \pm 0.106 $ & $0.000$                              & $0.406$ & $0.93$   &   $1072$                                          & $1066$ & $6 $  \\
Spitzer $[5.8]$ (Fig. \ref{fig:pl_spitzer_c_rich}c) & $10.789 \pm 0.040$ & $-7.689 \pm 0.047$ & $-7.183 \pm 0.130 $ & $0.000$                              & $0.494$ & $0.76$                &                  $1072$                         & $1066 $ & $6$   \\
Spitzer $[8.0]$ (Fig. \ref{fig:pl_spitzer_c_rich}d) & $10.691 \pm 0.044$ & $-7.786 \pm 0.050$ & $-8.329 \pm 0.143$ & $0.000$                              & $0.541$ & $0.96$               &             $1072$                              & $1065$ & $7$   \\
\hline
\hline
\multicolumn{10}{c}{O-rich Miras}       \\
\hline \hline

\multicolumn{10}{c}{linear fit$^a$, $\log P \leq 2.6$}       \\
\hline
WISE W1 (Fig. \ref{fig:pl_wise_o_rich_linear}a)         & $11.228 \pm 0.009 $ & $-7.249 \pm 0.025$ & $-3.807 \pm 0.066$ & $0.000$ & $ 0.116$ & $0.55$                & $196$                    & $194$ & $2 $  \\
WISE W2    (Fig. \ref{fig:pl_wise_o_rich_linear}b)       & $11.030 \pm 0.013$ & $-7.447 \pm 0.027$ & $-3.794 \pm 0.096$ & $0.000$ & $0.170$ & $0.80$                &          $196$                                 & $193$  & $2$   \\
WISE W3    (Fig. \ref{fig:pl_wise_o_rich_linear}c)        & $10.251 \pm 0.028$ & $-8.226 \pm 0.036$ & $-3.900 \pm 0.210$ & $0.000$ & $0.365$ & $0.94$                &   $199$                                        & $194$  & $5$    \\ \hline
Spitzer $[3.6]$ (Fig. \ref{fig:pl_spitzer_o_rich_linear}a) & $11.073 \pm 0.013$  & $-7.404 \pm 0.027$ & $-3.522 \pm 0.103$  & $0.000$ & $0.236$ & $0.73$                & $342$                      & $327$  & $15$   \\
Spitzer $[4.5]$ (Fig. \ref{fig:pl_spitzer_o_rich_linear}b) & $10.970 \pm 0.014$ & $-7.507 \pm 0.028$ & $-3.290 \pm 0.111$ & $0.000$ & $0.256$ & $0.78$               &     $342$                                      & $328$  & $14$    \\
Spitzer $[5.8]$ (Fig. \ref{fig:pl_spitzer_o_rich_linear}c)  & $10.792 \pm 0.014$ & $-7.685 \pm 0.028$ & $-3.313 \pm 0.111$ & $0.000$ & $0.254$ & $0.78$                &                                $342$           & $323$  & $19$    \\
Spitzer $[8.0]$ (Fig. \ref{fig:pl_spitzer_o_rich_linear}d) & $10.595 \pm 0.017$ & $-7.882 \pm 0.029 $ & $-3.473 \pm 0.131$ & $0.000$ & $0.303$ & $0.82$                &       $342$                                    & $330$  & $12$  \\
\hline

\multicolumn{10}{c}{quadratic fit$^b$}       \\
\hline
WISE W1 (Fig. \ref{fig:pl_wise_o_rich}a)         & $11.252 \pm 0.010 $ & $-7.224 \pm 0.026$ & $-3.749 \pm 0.075$ & $-1.810 \pm 0.154$ & $ 0.128$ & $0.92$                & $238$                     & $227$ & $11 $  \\
WISE W2    (Fig. \ref{fig:pl_wise_o_rich}b)       & $11.069 \pm 0.013$ & $-7.408 \pm 0.027$ & $-3.679 \pm 0.102$ & $-2.911 \pm 0.207$ & $0.173$ & $0.95$                &       $238$                                    & $231$  & $7$   \\
WISE W3    (Fig. \ref{fig:pl_wise_o_rich}c)        & $10.335 \pm 0.029$ & $-8.142 \pm 0.037$ & $-3.459 \pm 0.224$ & $-6.107 \pm 0.422$ & $0.383$ & $0.91$                &      $240$                                     & $232$  & $8$    \\ \hline
Spitzer $[3.6]$ (Fig. \ref{fig:pl_spitzer_o_rich}a) & $11.094 \pm 0.015$  & $-7.383 \pm 0.028$ & $-3.648 \pm 0.115$  & $-2.201 \pm 0.239$ & $0.267$ & $0.74$                & $398$                     & $381$  & $17$   \\
Spitzer $[4.5]$ (Fig. \ref{fig:pl_spitzer_o_rich}b) & $11.000 \pm 0.016$ & $-7.477 \pm 0.029$ & $-3.430 \pm 0.123$ & $-2.895 \pm 0.257$ & $0.287$ & $0.84$               &       $398$                                    & $385$  & $13$    \\
Spitzer $[5.8]$ (Fig. \ref{fig:pl_spitzer_o_rich}c)  & $10.813 \pm 0.017$ & $-7.664 \pm 0.029$ & $-3.568 \pm 0.128$ & $-2.736 \pm 0.256$ & $0.300$ & $0.81$                &   $398$                                        & $385$  & $13$    \\
Spitzer $[8.0]$ (Fig. \ref{fig:pl_spitzer_o_rich}d) & $10.640 \pm 0.018$ & $-7.837 \pm 0.030 $ & $-4.037 \pm 0.270$ & $-4.363 \pm 0.201$ & $0.315$ & $0.82$                &         $398$                                  & $385$  & $13$  \\
\hline
\label{tab:PL_results}
\end{tabular}
\end{center}
\footnotesize{$^a$ linear model in a form: $M_{\mathrm{\lambda, fit}} = a_{\mathrm{0, \lambda}} + a_{\mathrm{1,\lambda}} \times (\log{P}-2.3)$} \\
\footnotesize{$^b$ quadratic model in a form: $M_{\mathrm{\lambda, fit}} = a_{{0, \mathrm{\lambda}}} + a_{{\mathrm{1, \lambda}}} \times (\log{P}-2.3) + a_{{\mathrm{2, \lambda}}} \times (\log{P}-2.3)^2$} \\
\footnotesize{$^c$ scatter in the period--luminosity relation} \\
\footnotesize{$^d$ $\chi^2$ calculated using Equation \ref{eqn:matrix2} divided by the degrees of freedom}\\
\footnotesize{$^{\dagger}$ absolute zero-point with uncertainty calculated using distance modulus $\mu = 18.477$ mag \citep{2019Natur.567..200P}, and systematic and statistic uncertainties of the distance to the LMC}\\
\end{table*}

\section{Results} 
\label{sec:results}

\begin{figure*}
\gridline{\fig{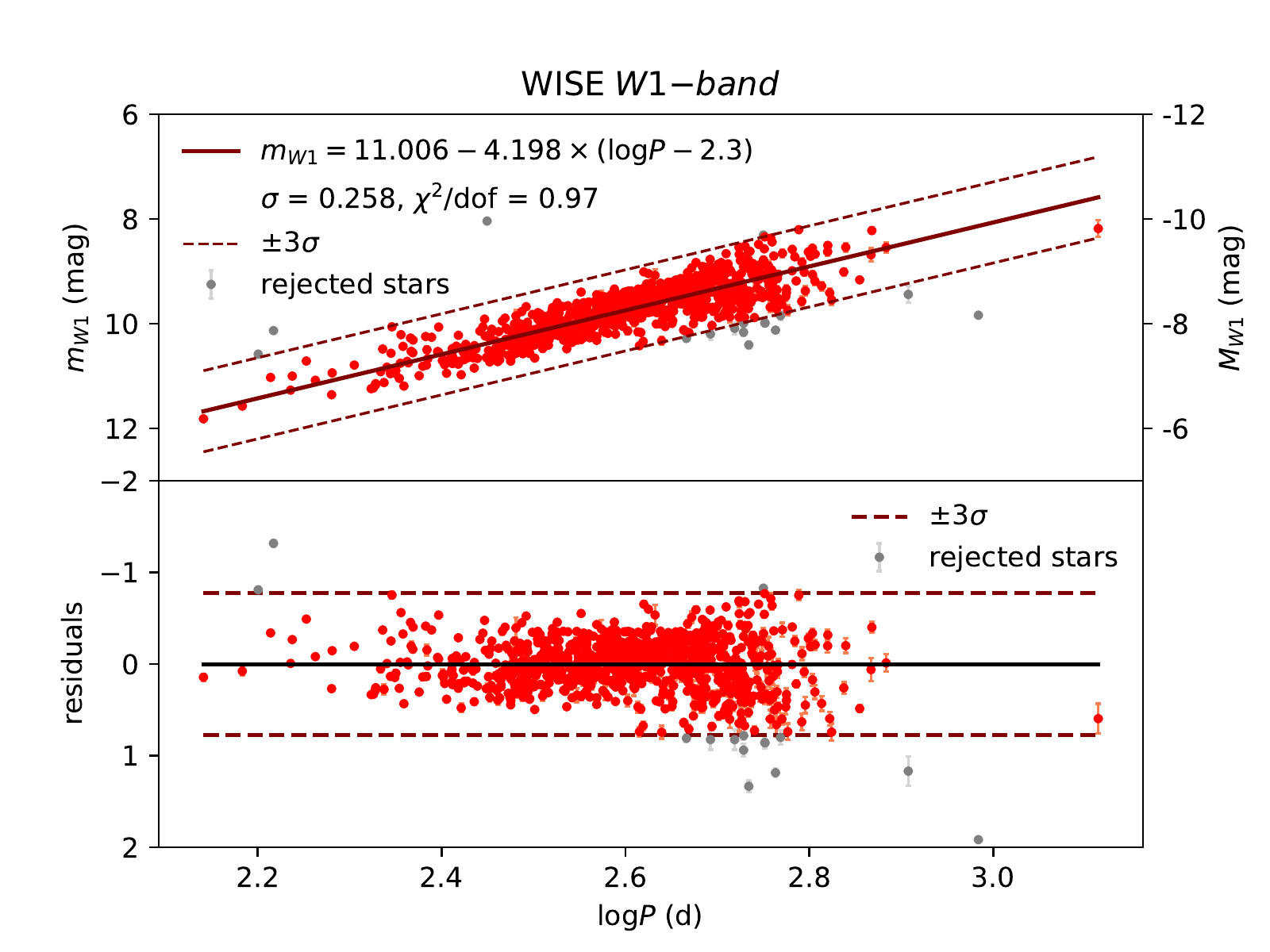}{0.5\textwidth}{(a)}
          \fig{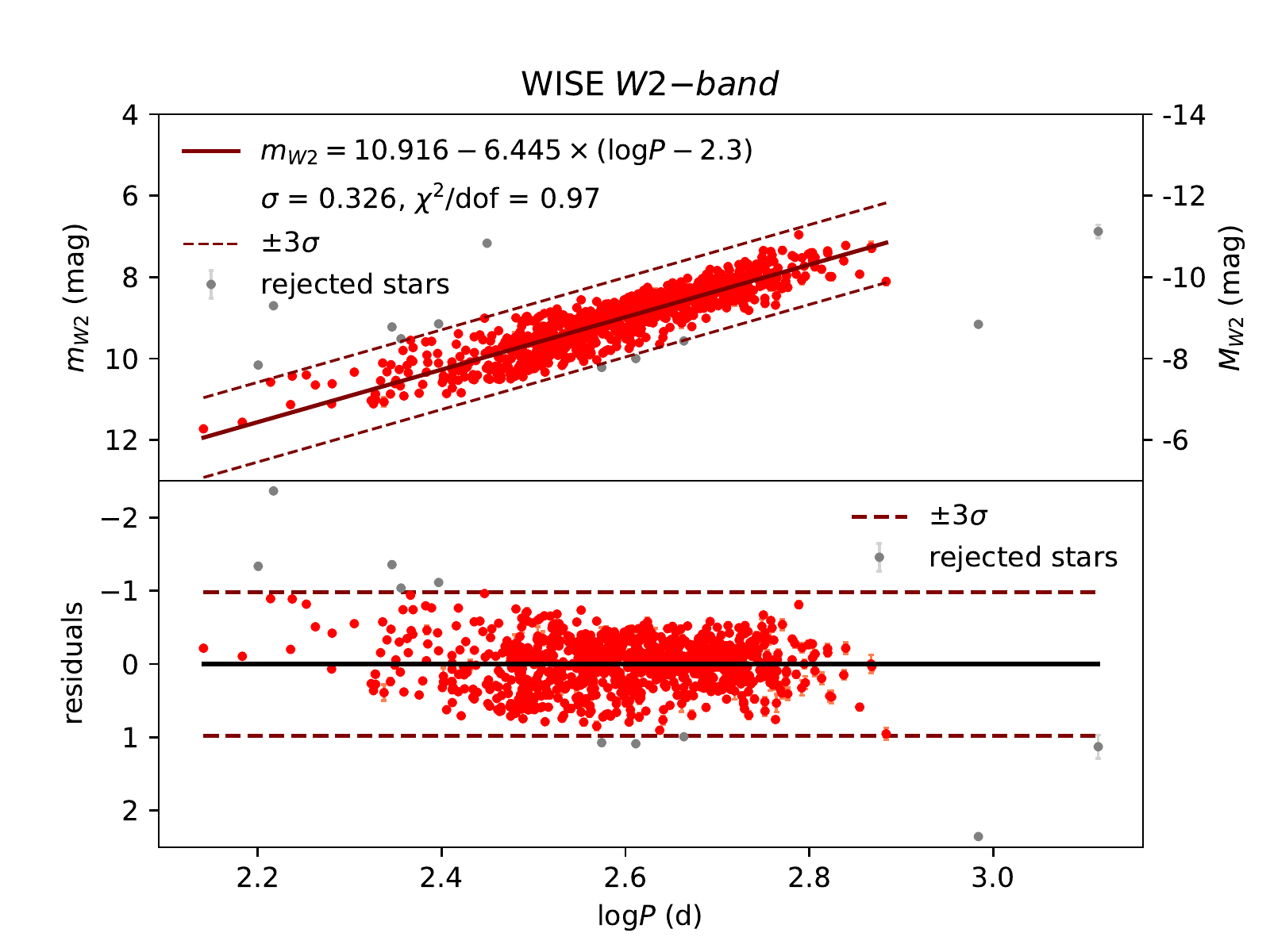}{0.5\textwidth}{(b)}
          }
\gridline{\fig{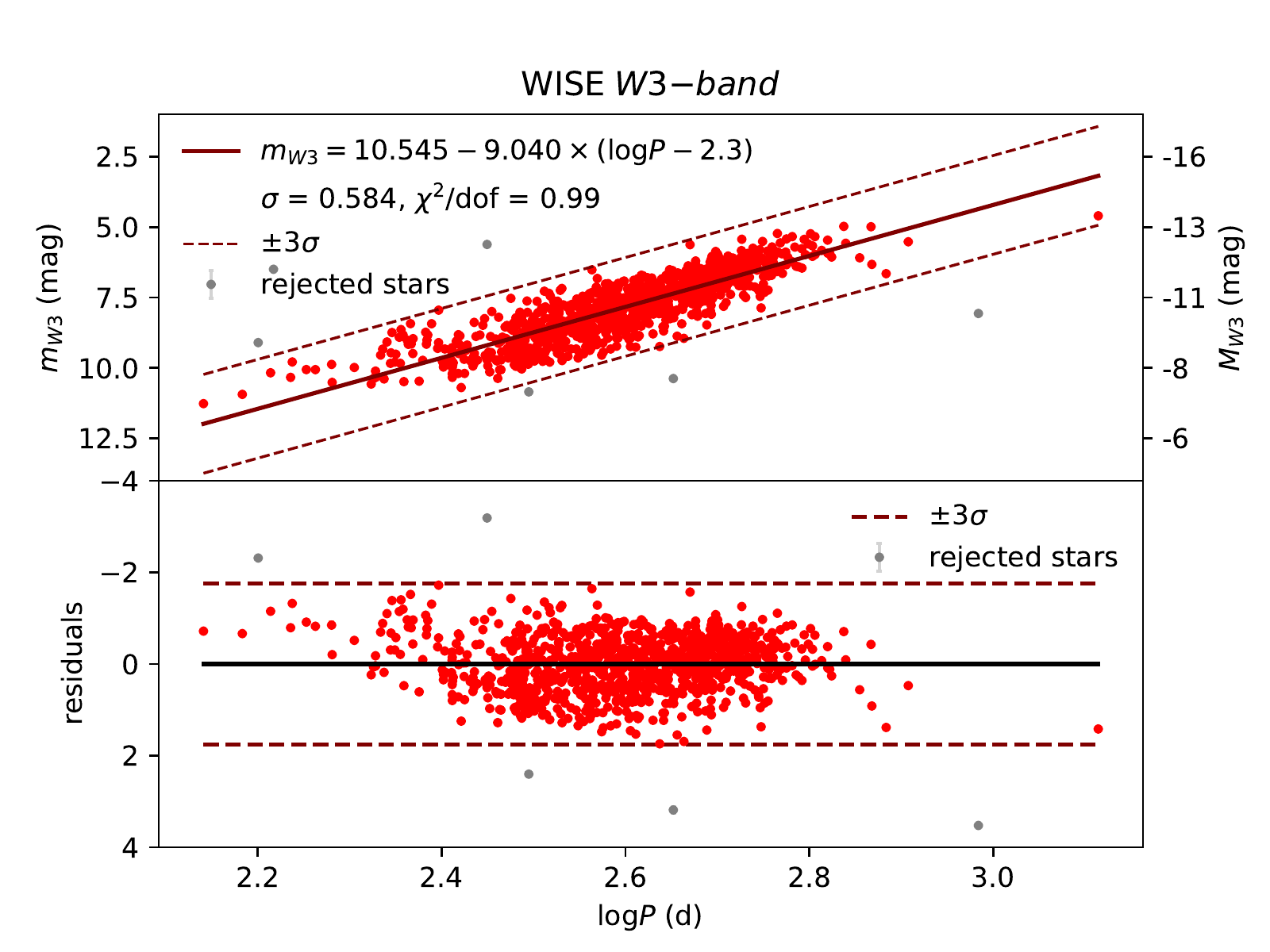}{0.5\textwidth}{(c)}
        \fig{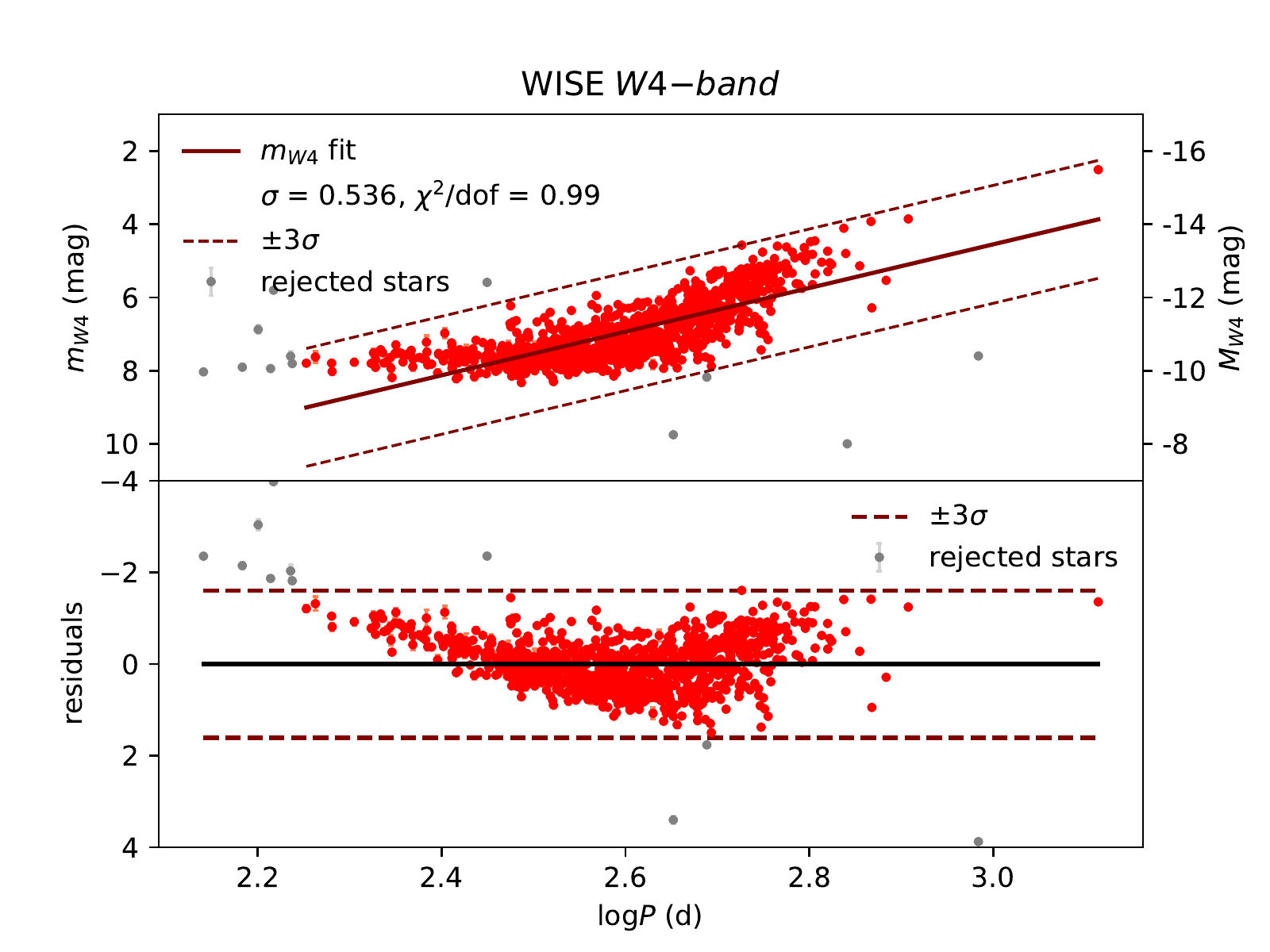}{0.5\textwidth}{(d)}
          }
\caption{Period--Luminosity Relations for C-rich Miras from the LMC in mid-IR WISE bands: (a) W1, (b) W2, (c) W3, (d) W4. In each case, the left (right) y-axis of the top panel presents the observed (absolute) mean magnitude in a given band $m_{\mathrm{{\lambda}}}$ as a function of the logarithm of the pulsation period $P$ (in days), while the bottom panel presents residuals calculated as the difference between the measured and predicted absolute magnitudes from the fit. In the top panel, we present the best fit with the solid line, while dashed lines present the range of $\pm 3\sigma$, where $\sigma$ means the dispersion along the fit. The model is given in each plot (with the exception of panel (d) for the $W4$-band due to the photometric issues in this band) in the form: $m_{\mathrm{\lambda, fit}} = a_{\mathrm{0, \lambda}} + a_{\mathrm{1,\lambda}} \times (\log{P}-2.3)$. Final samples of C-rich Miras are plotted with red points. Errorbars of $\log{P}$ are smaller than the point size. Rejected stars during $3\sigma$-clipping procedure are plotted in gray. The final $3\sigma$ range, above which there are no outliers, is plotted as dashed line in the bottom plot. Panel (d) for the $W4-$band is shown here for completeness.}
\label{fig:pl_wise_c_rich}
\end{figure*}

\begin{figure*}
\gridline{\fig{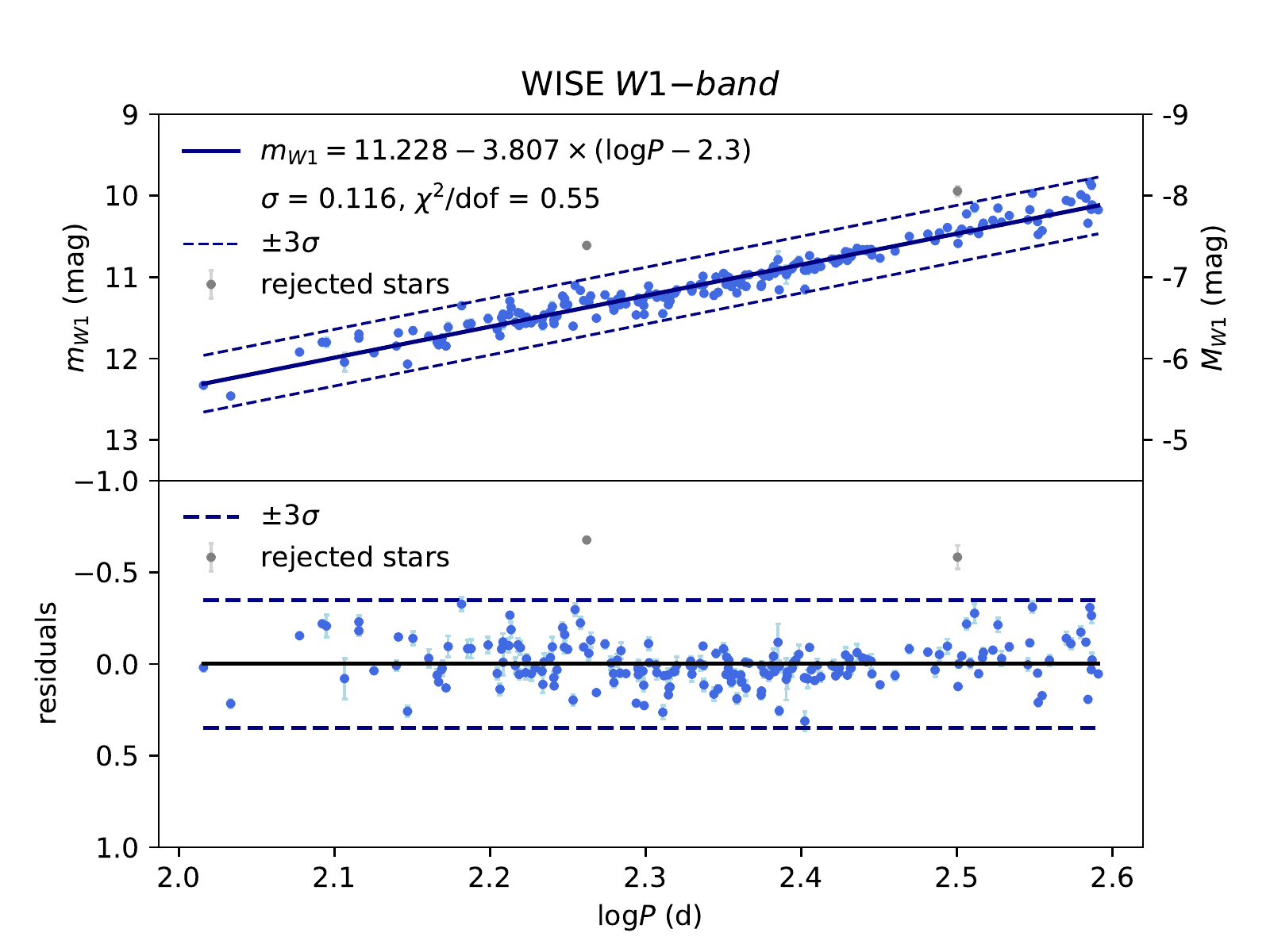}{0.5\textwidth}{(a)}
          \fig{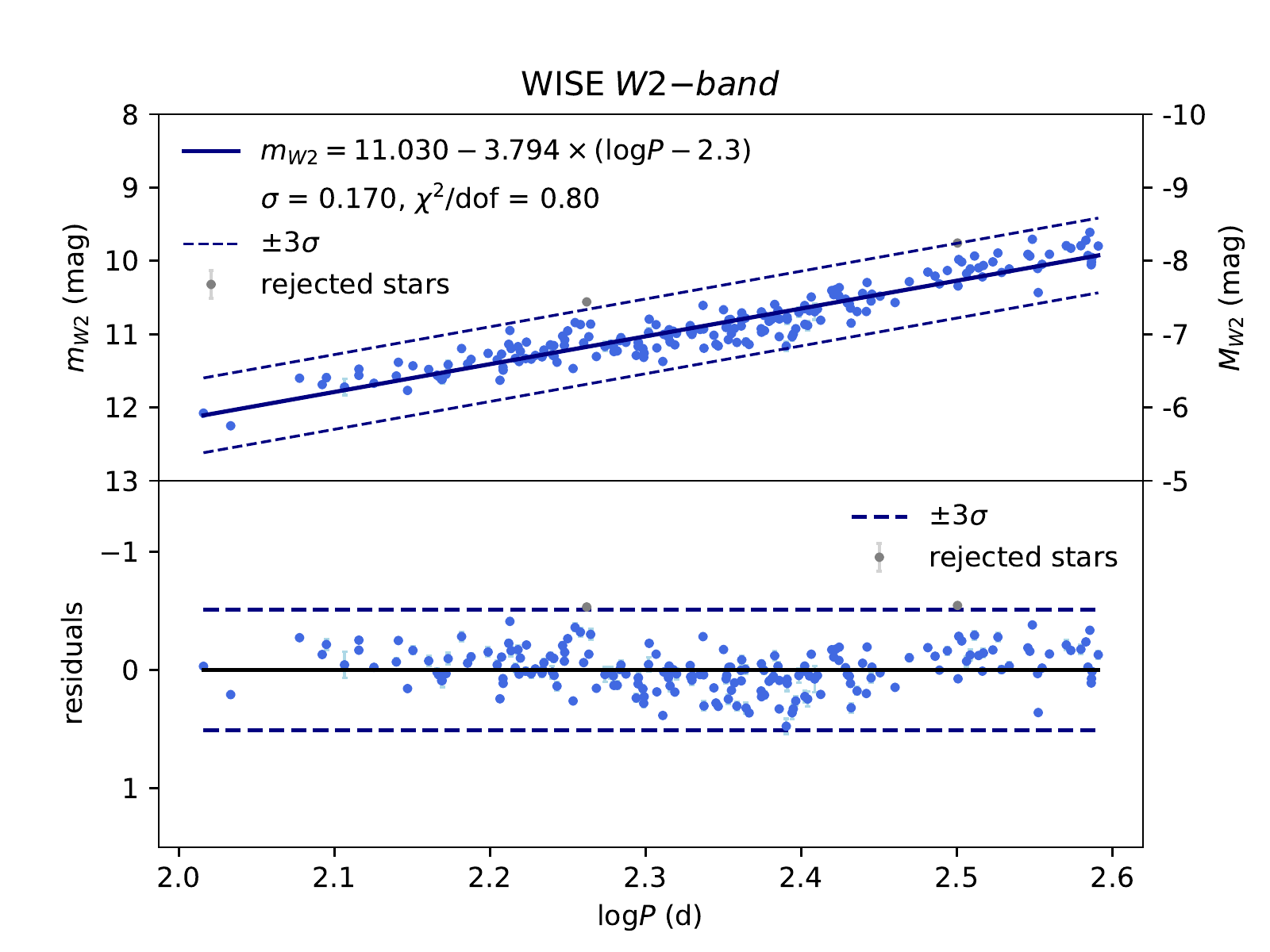}{0.5\textwidth}{(b)}
          }
\gridline{\fig{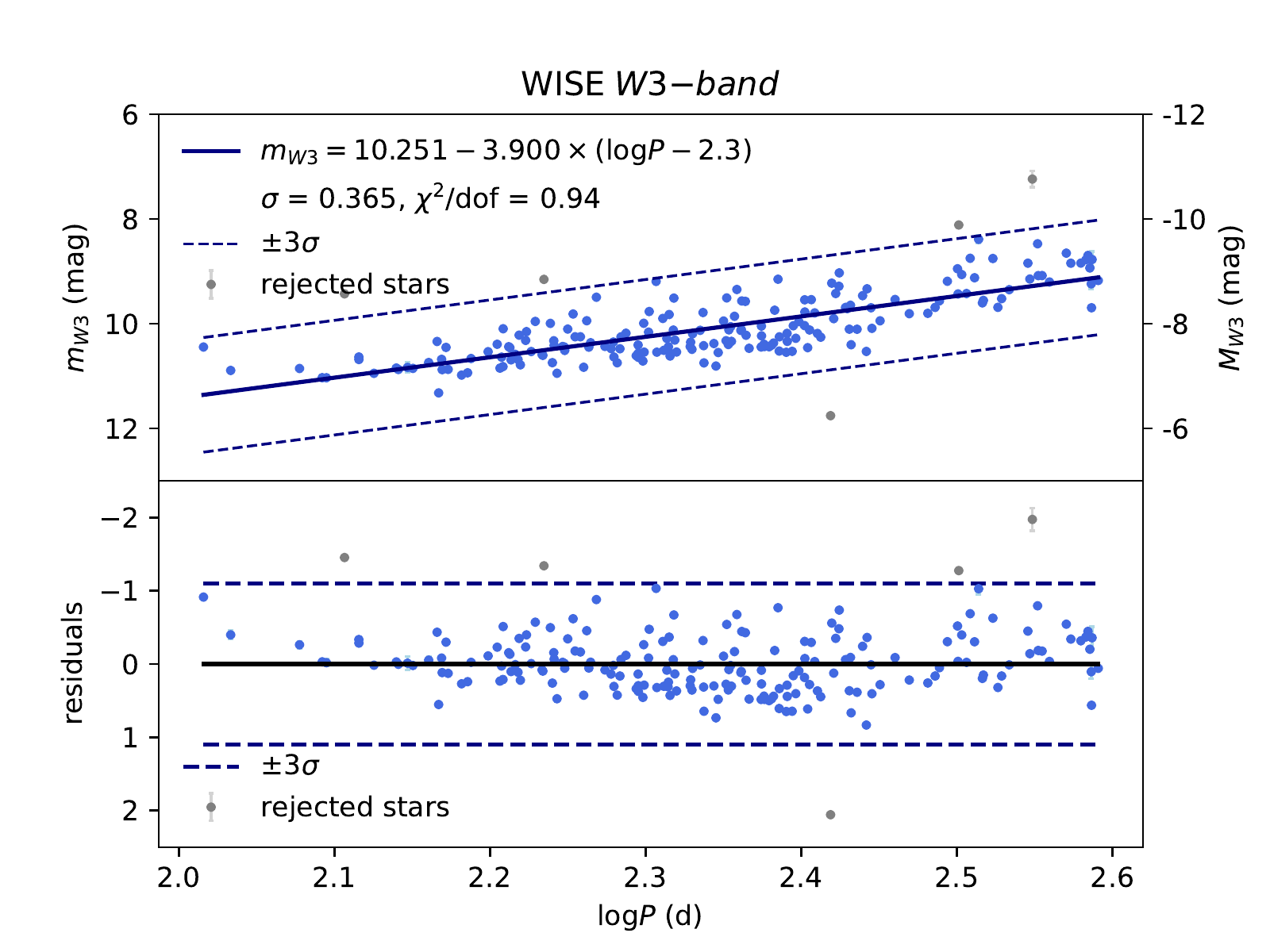}{0.5\textwidth}{(c)}
            \fig{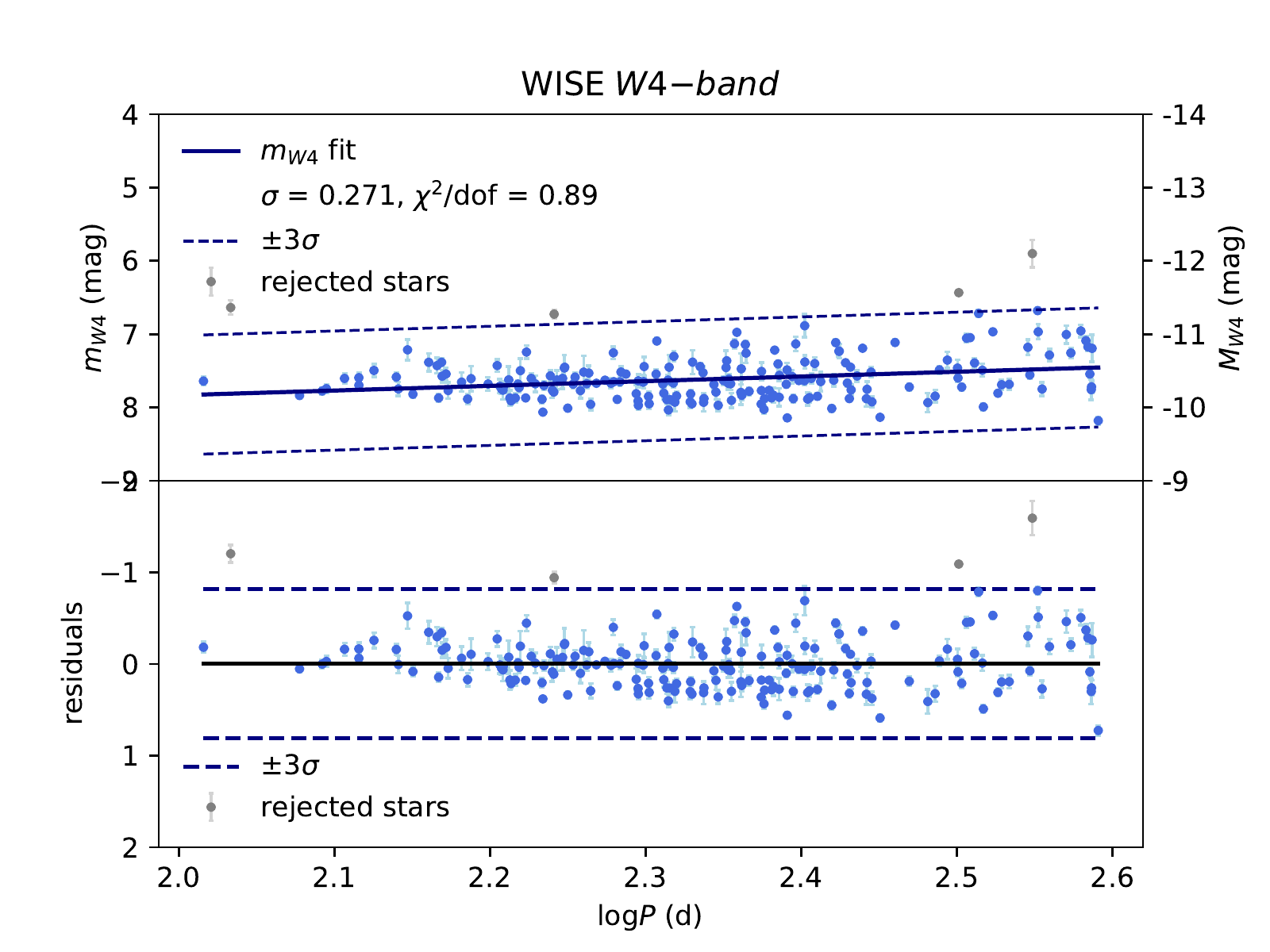}{0.5\textwidth}{(d)}
          }
\caption{Period--Luminosity Relations for O-rich Miras from the LMC in mid-IR WISE bands: (a) W1, (b) W2, (c) W3, (d) W4. In each case, the left (right) y-axis of the top panel presents the observed (absolute) mean magnitude in a given band $m_{\mathrm{{\lambda}}}$ as a function of the logarithm of the pulsation period $P$ (in days), while the bottom panel presents residuals calculated as the difference between the measured and predicted absolute magnitudes from the fit. In the top panel, we present the best fit with the solid line, while dashed lines present the range of $\pm 3\sigma$, where $\sigma$ means the dispersion along the fit. The model is given in each plot (with the exception of panel (d) for $W4$-band due to the photometric issues in this band) in the form: $m_{\mathrm{\lambda, fit}} = a_{{0, \mathrm{\lambda}}} + a_{{\mathrm{1, \lambda}}} \times (\log{P}-2.3)$. Final samples of O-rich Miras are plotted with blue points. Errorbars of $\log{P}$ are smaller than the point size. Rejected stars during $3\sigma$-clipping procedure are plotted in gray. The final $3\sigma$ range, above which there are no outliers, is plotted as dashed line in the bottom plot. Panel (d) for the $W4-$band is shown here for completeness.}
\label{fig:pl_wise_o_rich_linear}
\end{figure*}

\begin{figure*}
\gridline{\fig{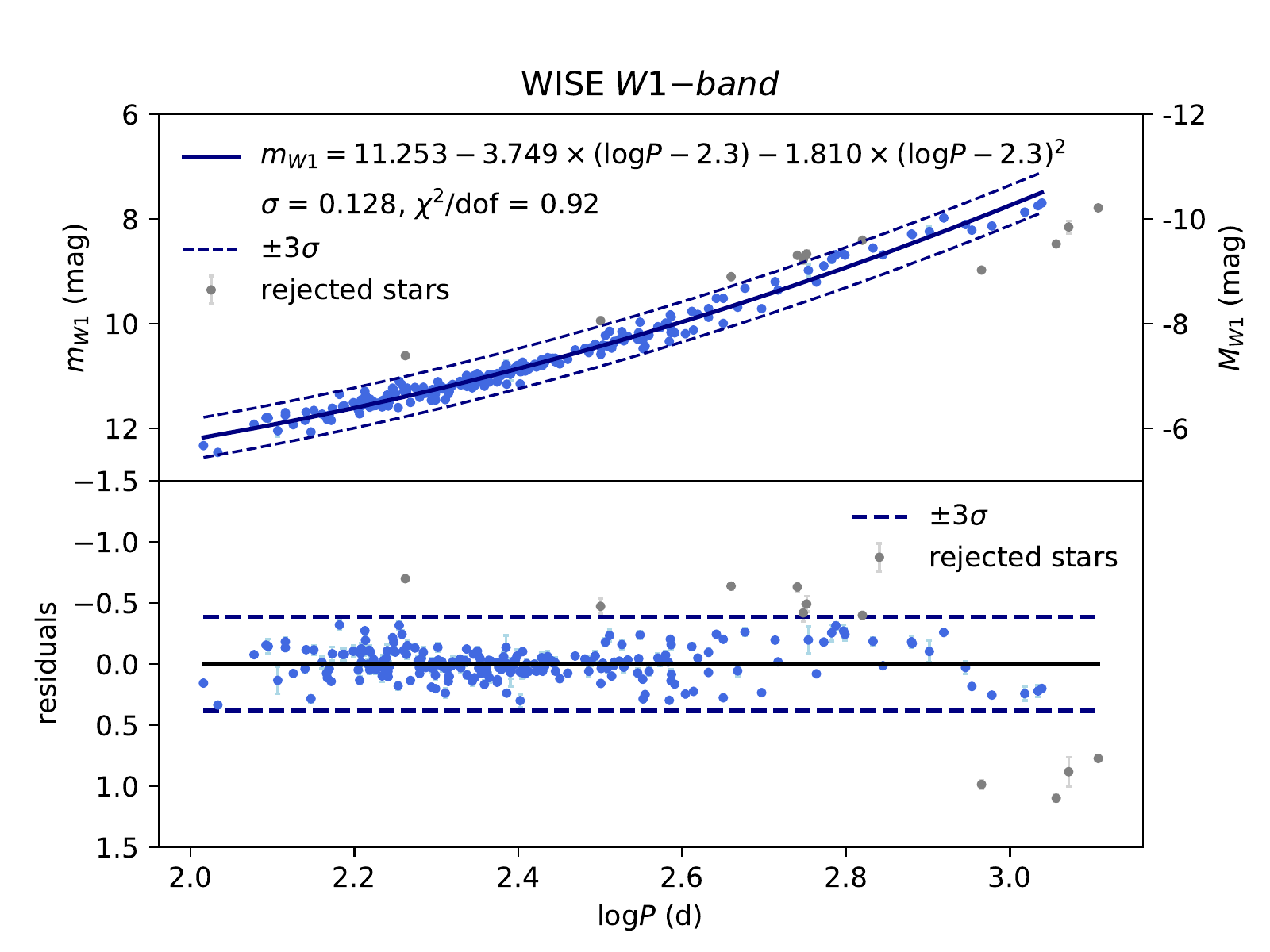}{0.5\textwidth}{(a)}
          \fig{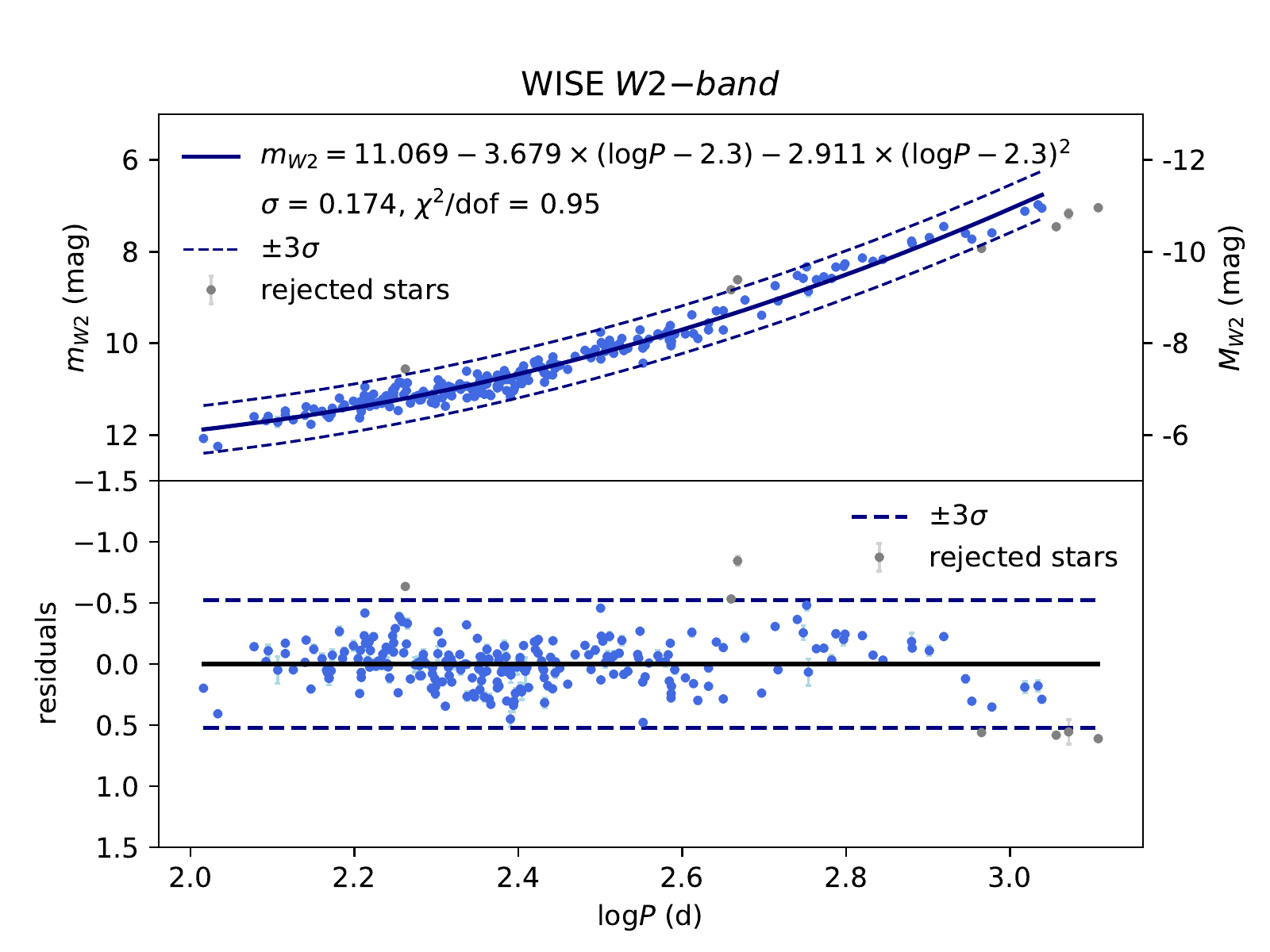}{0.5\textwidth}{(b)}
          }
\gridline{\fig{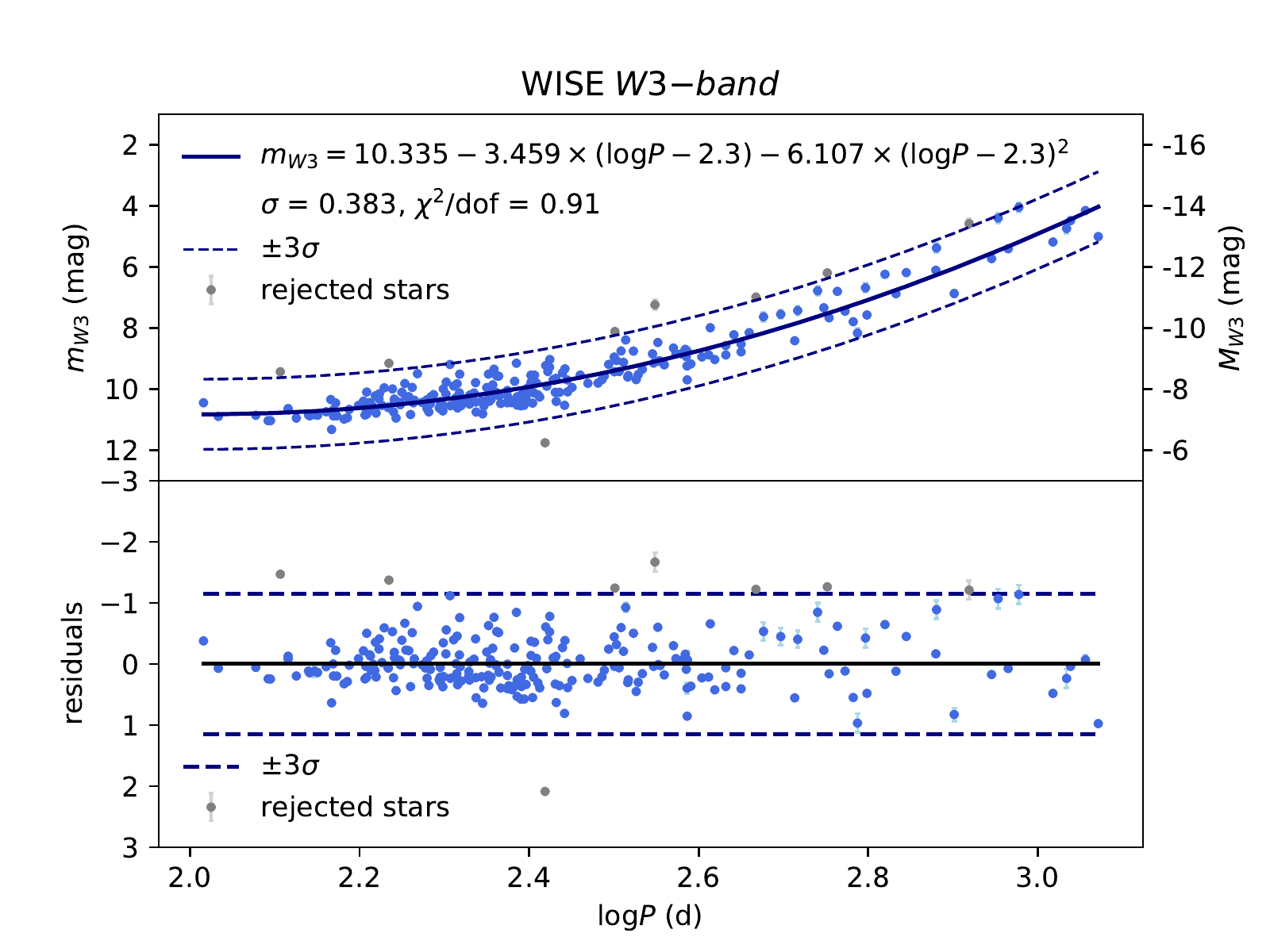}{0.5\textwidth}{(c)}
            \fig{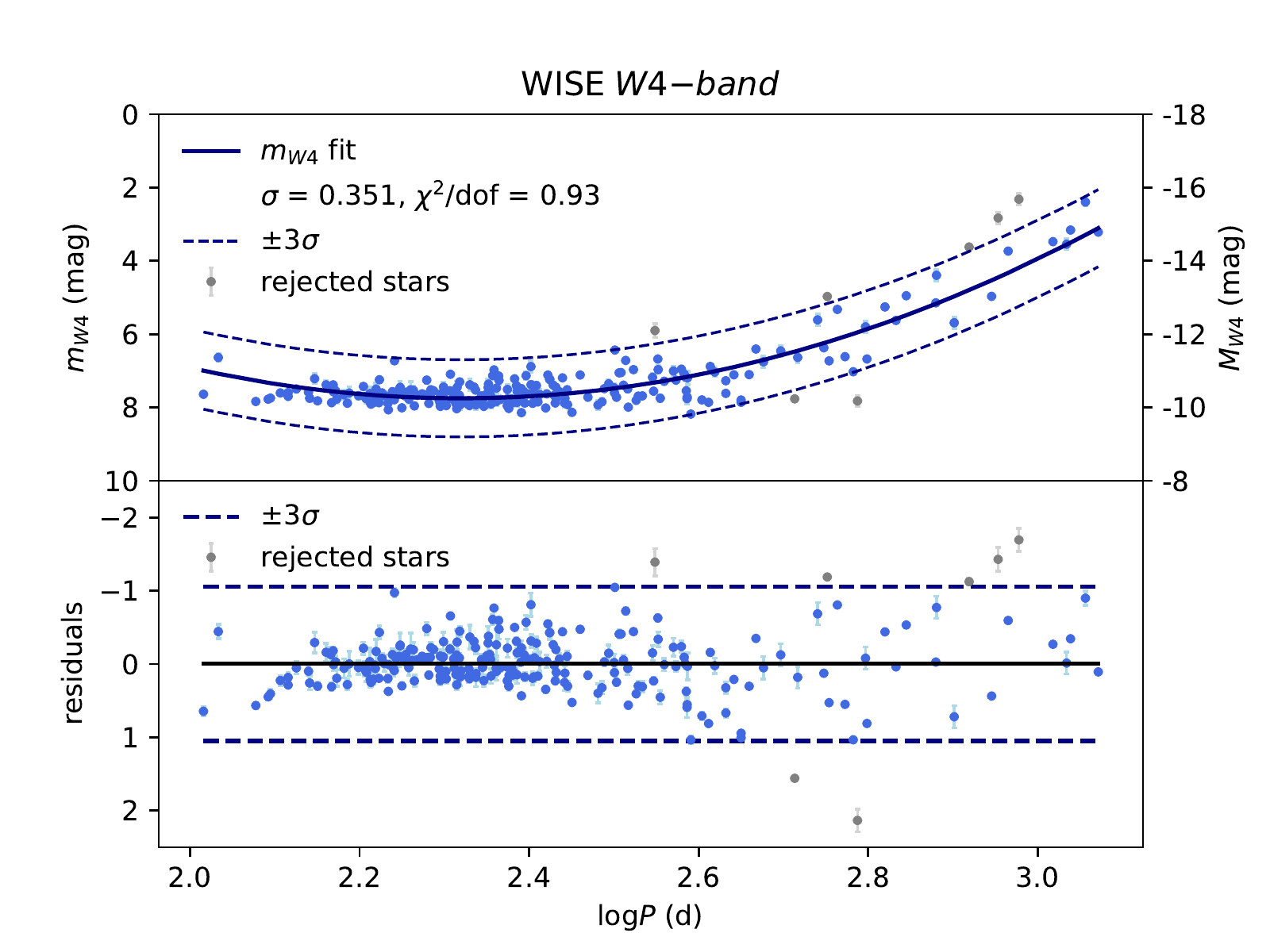}{0.5\textwidth}{(d)}
          }
\caption{Period--Luminosity Relations for O-rich Miras from the Large Magellanic Cloud in mid-IR WISE bands: (a) W1, (b) W2, (c) W3, (d) W4. In each case, the left (right) y-axis of the top panel presents the observed (absolute) mean magnitude in a given band $m_{\mathrm{{\lambda}}}$ as a function of the logarithm of the pulsation period $P$ (in days), while the bottom panel presents residuals calculated as the difference between the measured and predicted absolute magnitudes from the fit. In the top panel, we present the best fit with the solid line, while dashed lines present the range of $\pm 3\sigma$, where $\sigma$ means the dispersion along the fit. The model is given in each plot (with the exception of panel (d) for $W4$-band due to the photometric issues in this band) in the form: $m_{\mathrm{\lambda, fit}} = a_{{0, \mathrm{\lambda}}} + a_{{\mathrm{1, \lambda}}} \times (\log{P}-2.3) + a_{{\mathrm{2, \lambda}}} \times (\log{P}-2.3)^2$. Final samples of O-rich Miras are plotted with blue points. Errorbars of $\log{P}$ are smaller than the point size. Rejected stars during $3\sigma$-clipping procedure are plotted in gray. The final $3\sigma$ range, above which there are no outliers, is plotted as dashed line in the bottom plot. Panel (d) for the $W4-$band is shown here for completeness.}
\label{fig:pl_wise_o_rich}
\end{figure*}

Using the PLR models and fitting procedures, described in Section \ref{subsec:models_determining}, we performed fits to the datasets using the weighted least squares method with a $3\sigma$-clipping procedure. We iteratively fitted the linear model in a form of Equation~(\ref{eqn:linear_fit}) to the O-rich and C-rich Miras, and the quadratic model in a form of Equation~(\ref{eqn:quadratic_fit}) to the O-rich Miras only. In each iteration, we rejected points deviating by more than $3\sigma$ from the model, until no outliers were left. The final parameters of the fitted models (mid-IR PLRs), the absolute zero-point $a_\mathrm{0, \lambda, abs}$ with its uncertainty, are presented in Table \ref{tab:PL_results}, along with dispersions $\sigma_{\mathrm{\lambda}}$, $\chi^2/\mathrm{dof}$, the number of initial data points $N_{\mathrm{in}}$, the final number of data points after rejecting outliers $N_{\mathrm{fin}}$, and the number of outliers rejected during $3\sigma$-clipping procedure $N_{\mathrm{out}}$. PLRs for each band are presented in Figures: \ref{fig:pl_wise_c_rich} (WISE, C-rich Miras), \ref{fig:pl_wise_o_rich_linear} (WISE, O-rich Miras, linear fit), \ref{fig:pl_wise_o_rich} (WISE, O-rich Miras, quadratic fit), \ref{fig:pl_spitzer_c_rich} (Spitzer, \mbox{C-rich} Miras), \ref{fig:pl_spitzer_o_rich_linear} (Spitzer, O-rich Miras, linear fit), \ref{fig:pl_spitzer_o_rich} (Spitzer, O-rich Miras, quadratic fit).

As mentioned in Section \ref{subsec:mid-IR}, the SEDs fitting showed that photometric measurements in the $W4$-band are significantly biased. Therefore, in Figures \ref{fig:pl_wise_c_rich}, \ref{fig:pl_wise_o_rich_linear}, and \ref{fig:pl_wise_o_rich}, we present the PLRs in $W4-$band for completeness purposes only. We do not provide the fit parameters in Table \ref{tab:PL_results} and in Figures \ref{fig:pl_wise_c_rich}, \ref{fig:pl_wise_o_rich_linear}, and \ref{fig:pl_wise_o_rich}, because PLRs based on biased photometric magnitudes lead to biased PLRs, and should not be used in future studies.

PLRs for the C-rich Miras, both in the WISE and Spitzer bands, have significantly larger dispersions $\sigma_\mathrm{\lambda}$ than the ones obtained for the O-rich Mira PLRs. One of the reasons for that is the presence of gas and dust ejecta from these stars that can significantly alter their luminosity (by introducing long-term trends or variations), which in turn can lead to larger dispersions. As a rule of thumb, changes of brightness (the mean and amplitude) of O-rich Miras are reasonably stable over time, so that scatter along PLRs is smaller than for the C-rich Miras.

\subsection{Comparison of mean magnitude measuring methods} \label{sec:mean_mag_comp}

One of the basic parameters that characterizes a star is their mean magnitude. The mean magnitude should not depend on the measurement method, however, various methods come with their own advantages and disadvantages, the latter leading to biases. Since the mean magnitudes are extremely important for the PLRs analyses, we investigated the influence of several methods on the PLRs zero-point $a_{\mathrm{0, \lambda}}$, slope $a_{\mathrm{1, \lambda}}$, and dispersion $\sigma_\lambda$.

\begin{figure*}
\gridline{\fig{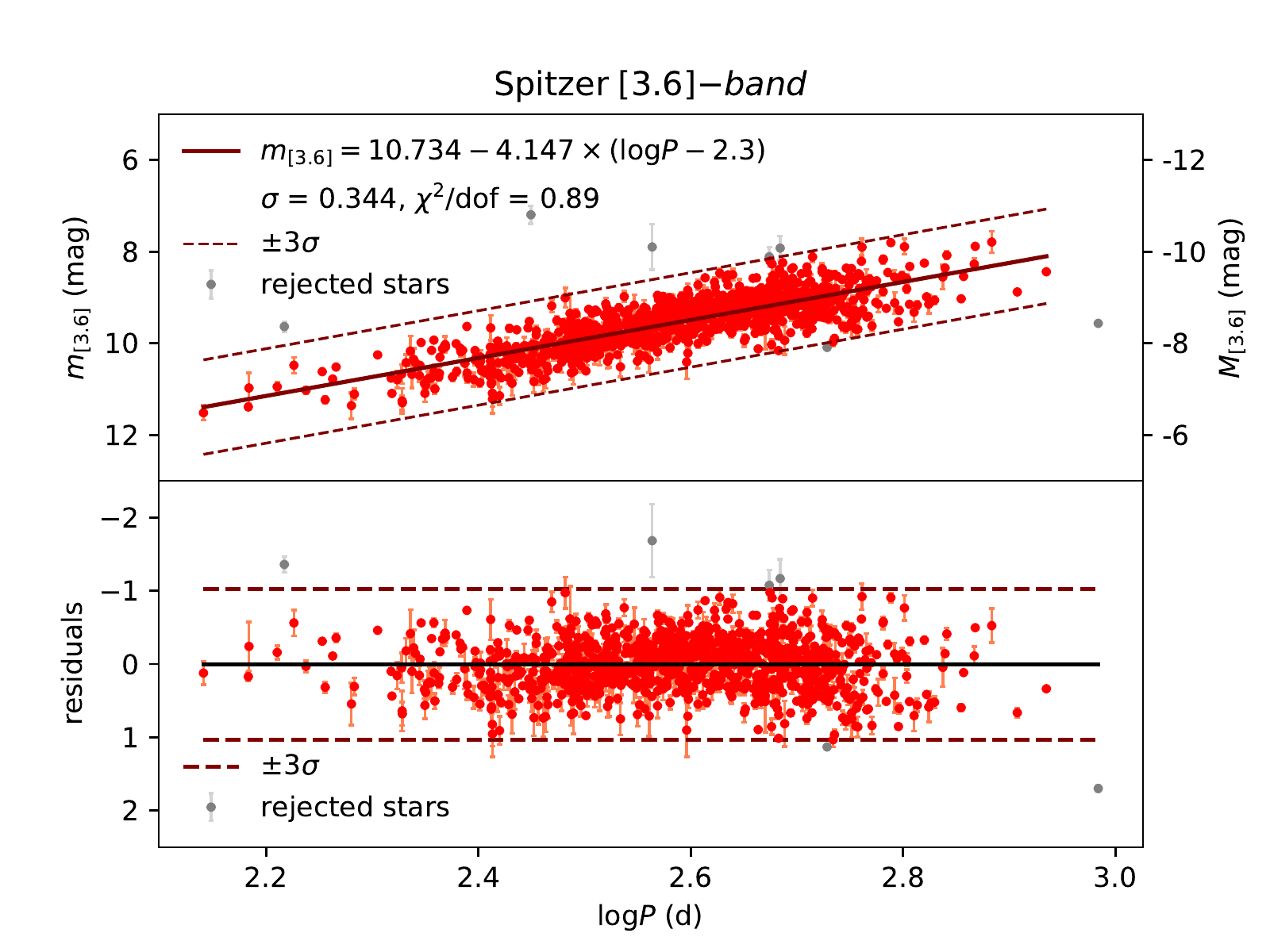}{0.5\textwidth}{(a)}
          \fig{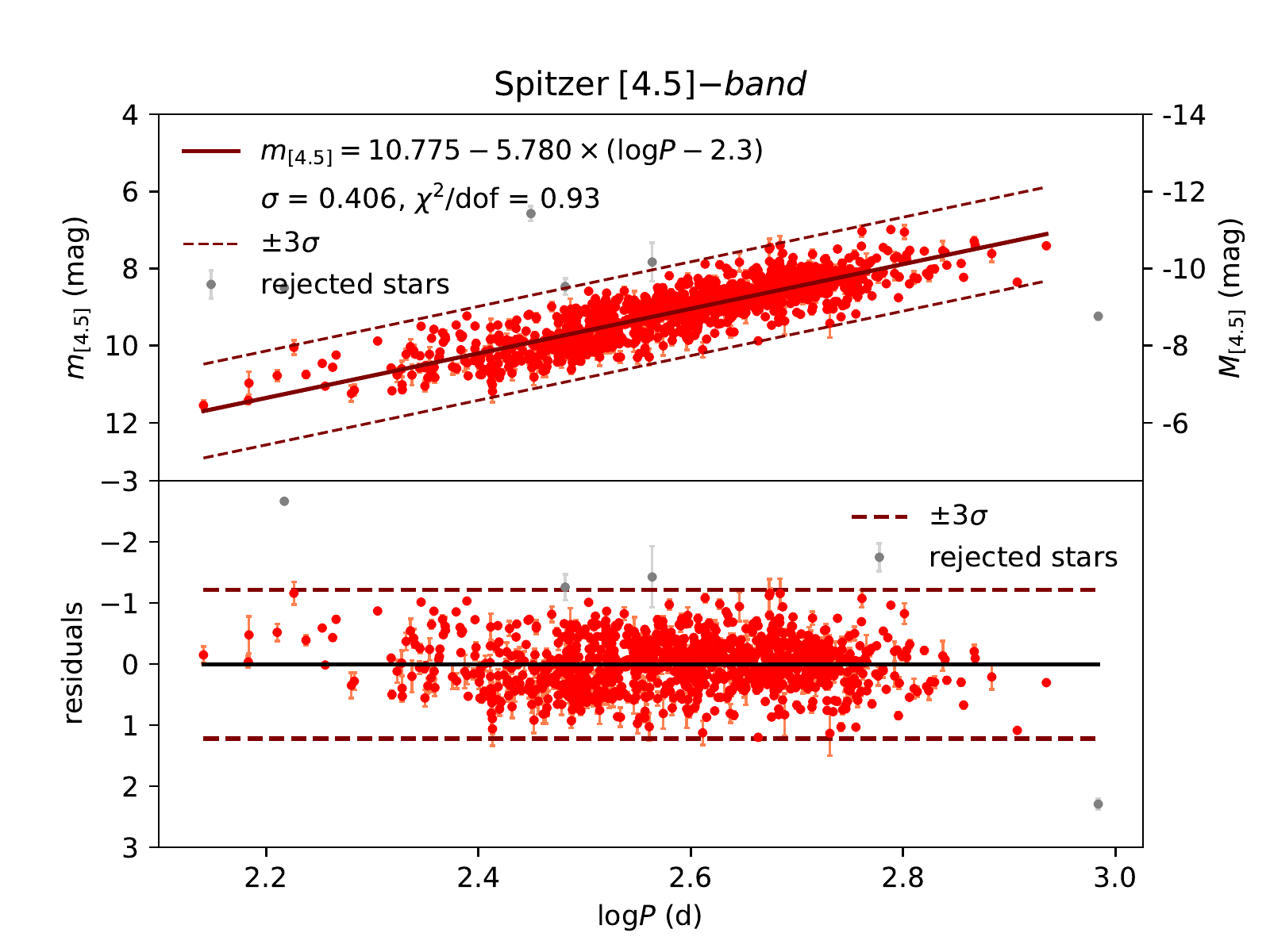}{0.5\textwidth}{(b)}
          }
\gridline{\fig{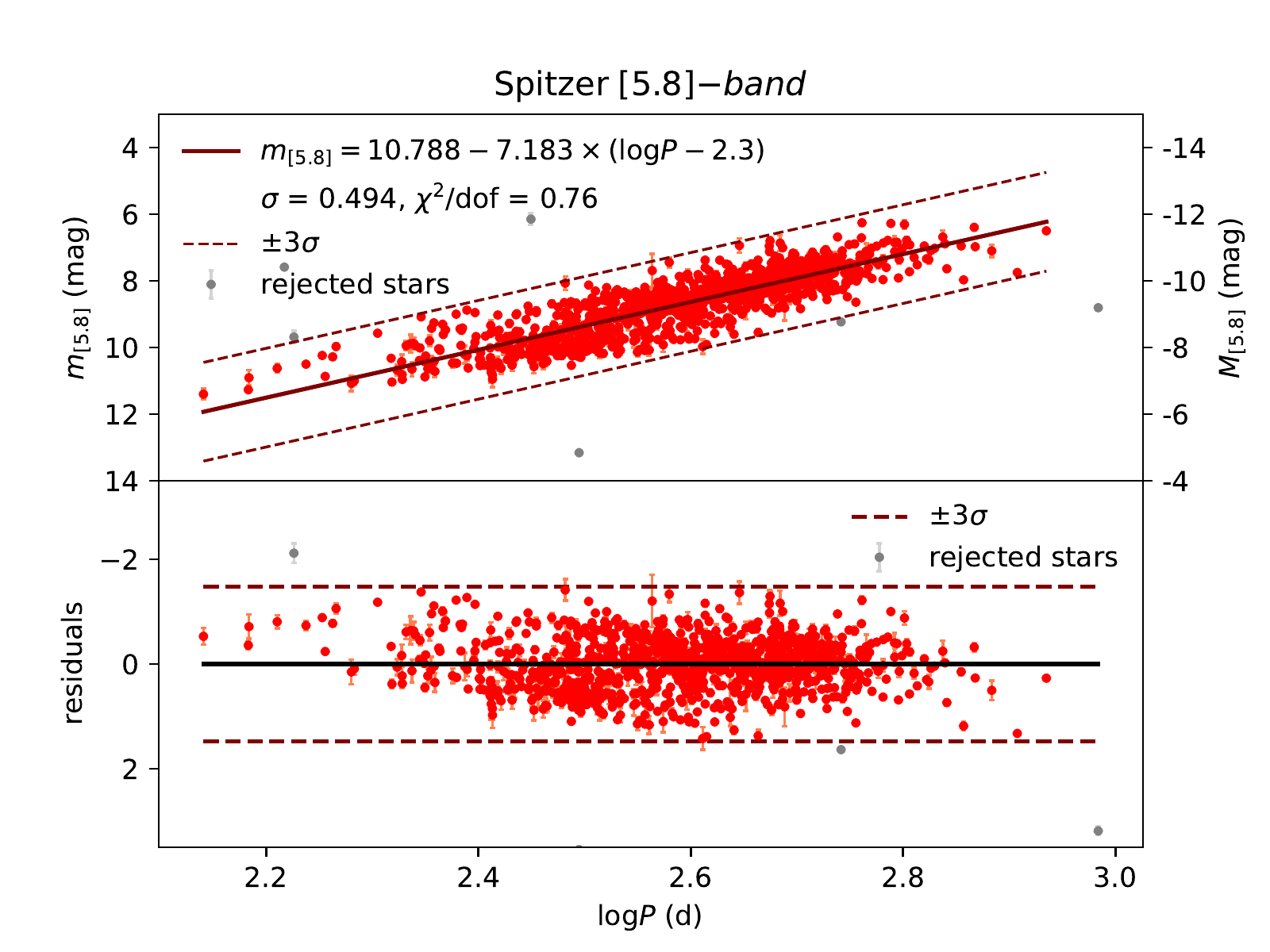}{0.5\textwidth}{(c)}
          \fig{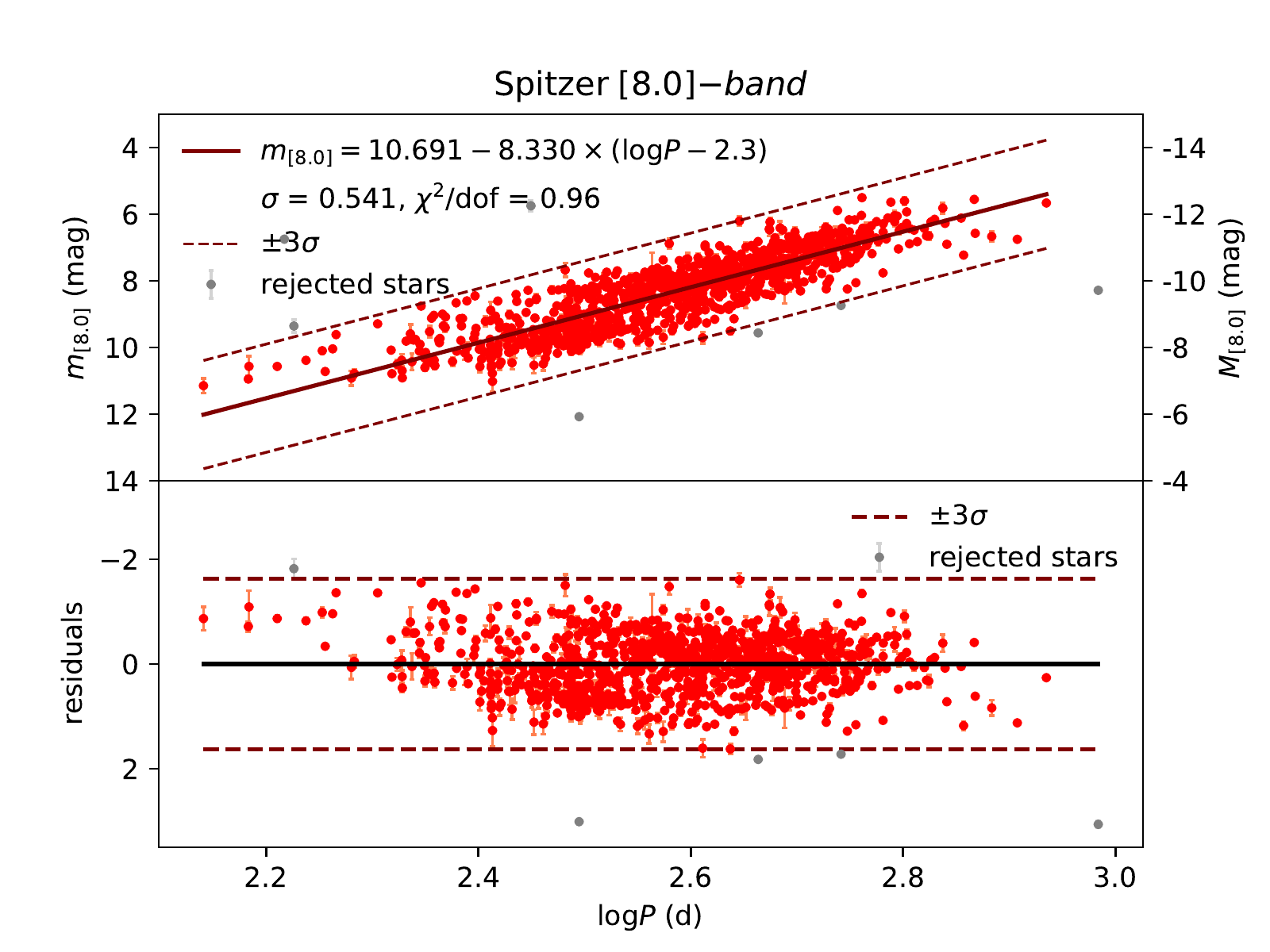}{0.5\textwidth}{(d)}
          }
\caption{Same as in Figure \ref{fig:pl_wise_c_rich}, but for the Spitzer bands: (a) $[3.6] \; \mu m$, (b) $[4.5] \; \mu m$, (c) $[5.8] \; \mu m$, (d) $[8.0] \; \mu m$.}
\label{fig:pl_spitzer_c_rich}
\end{figure*}

\begin{figure*}
\gridline{\fig{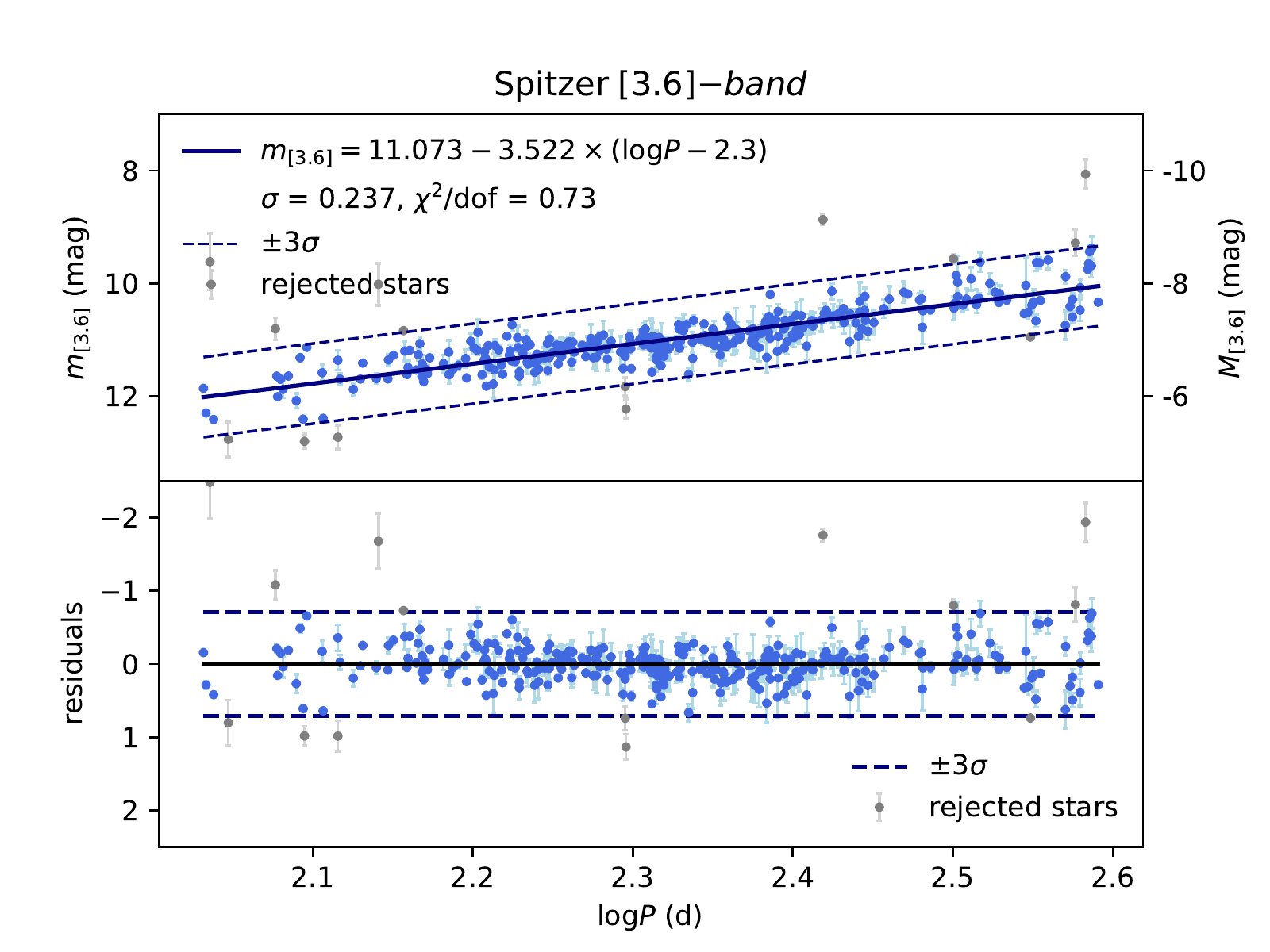}{0.5\textwidth}{(a)}
          \fig{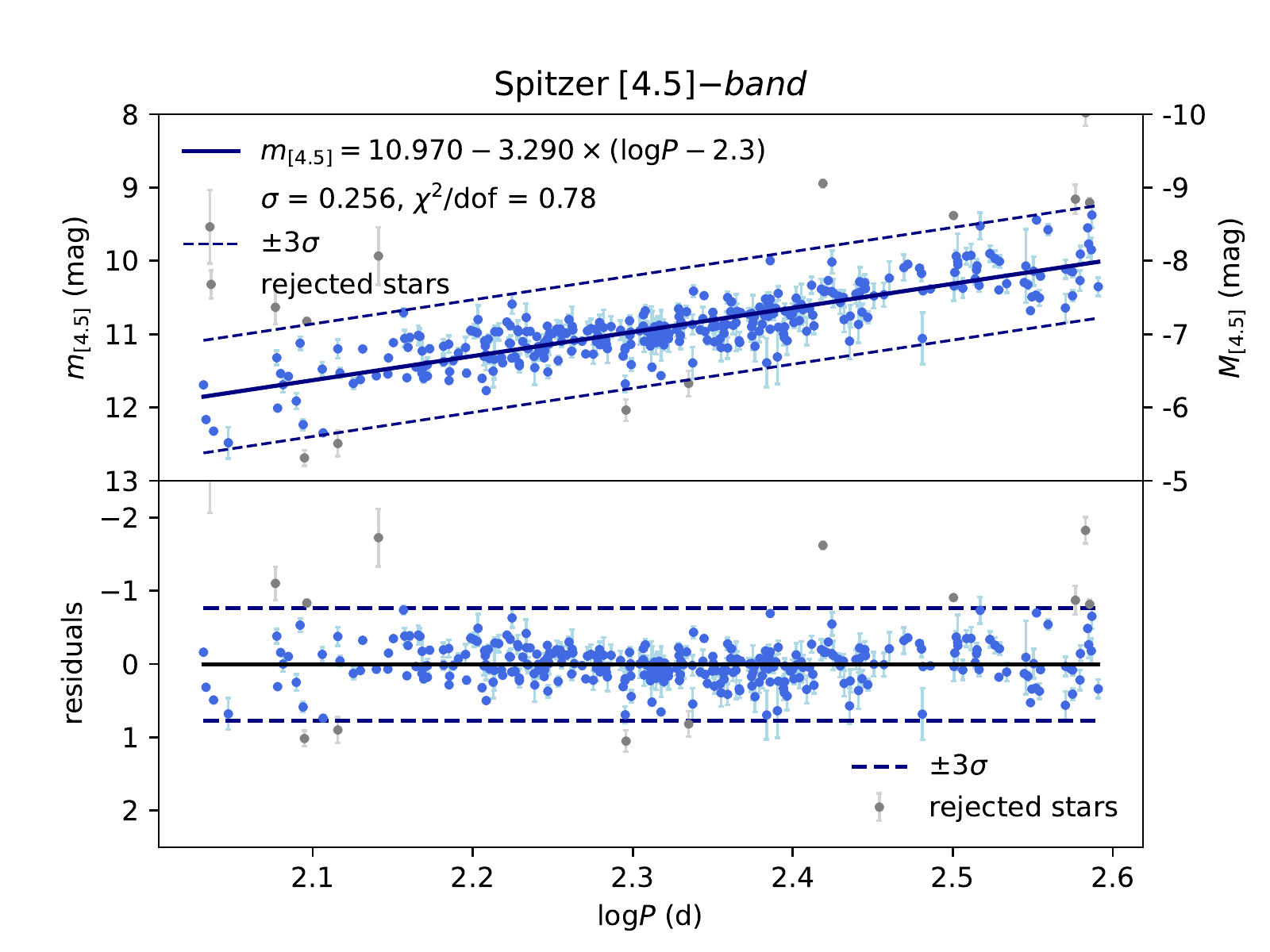}{0.5\textwidth}{(b)}
          }
\gridline{\fig{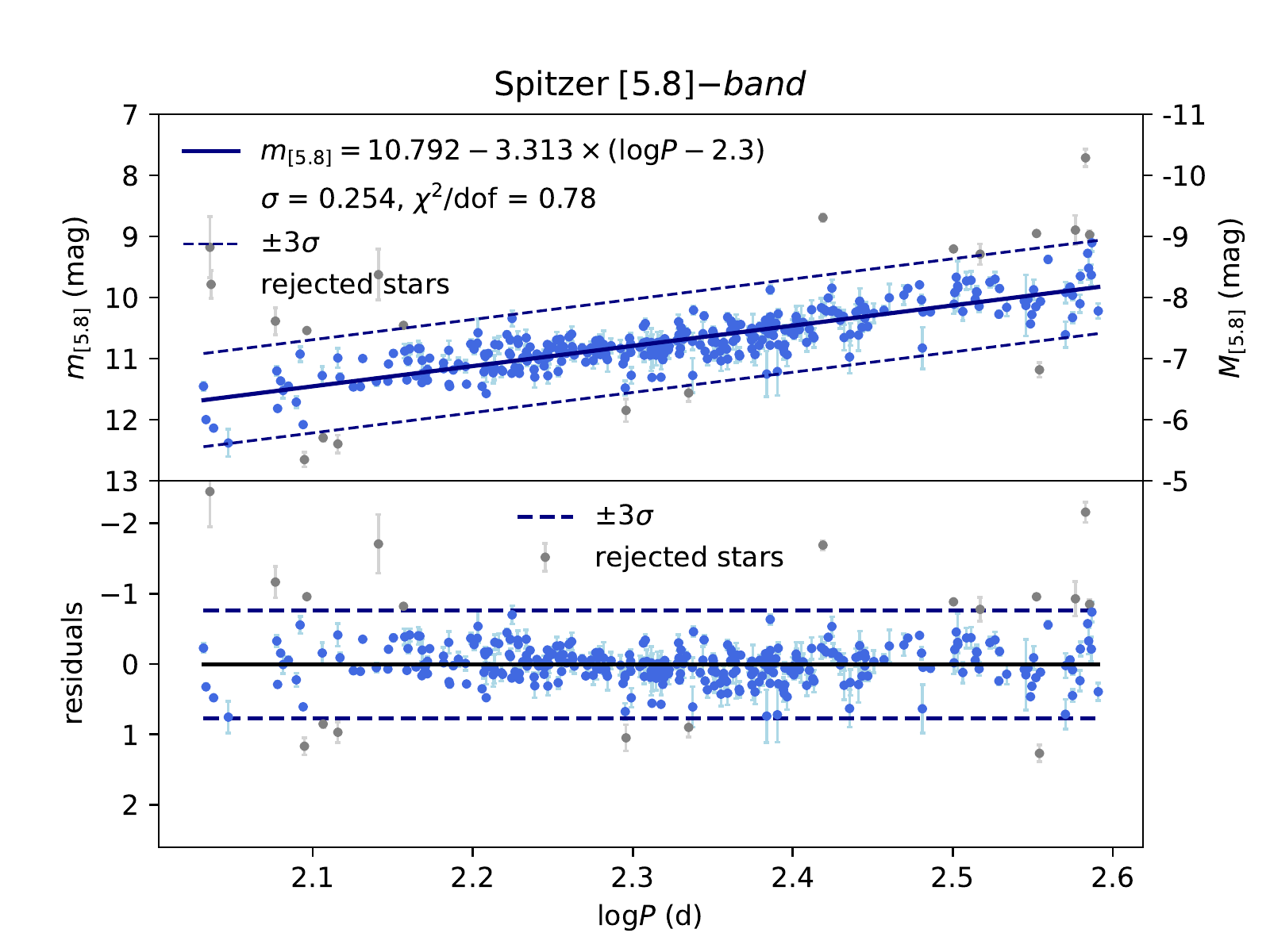}{0.5\textwidth}{(c)}
          \fig{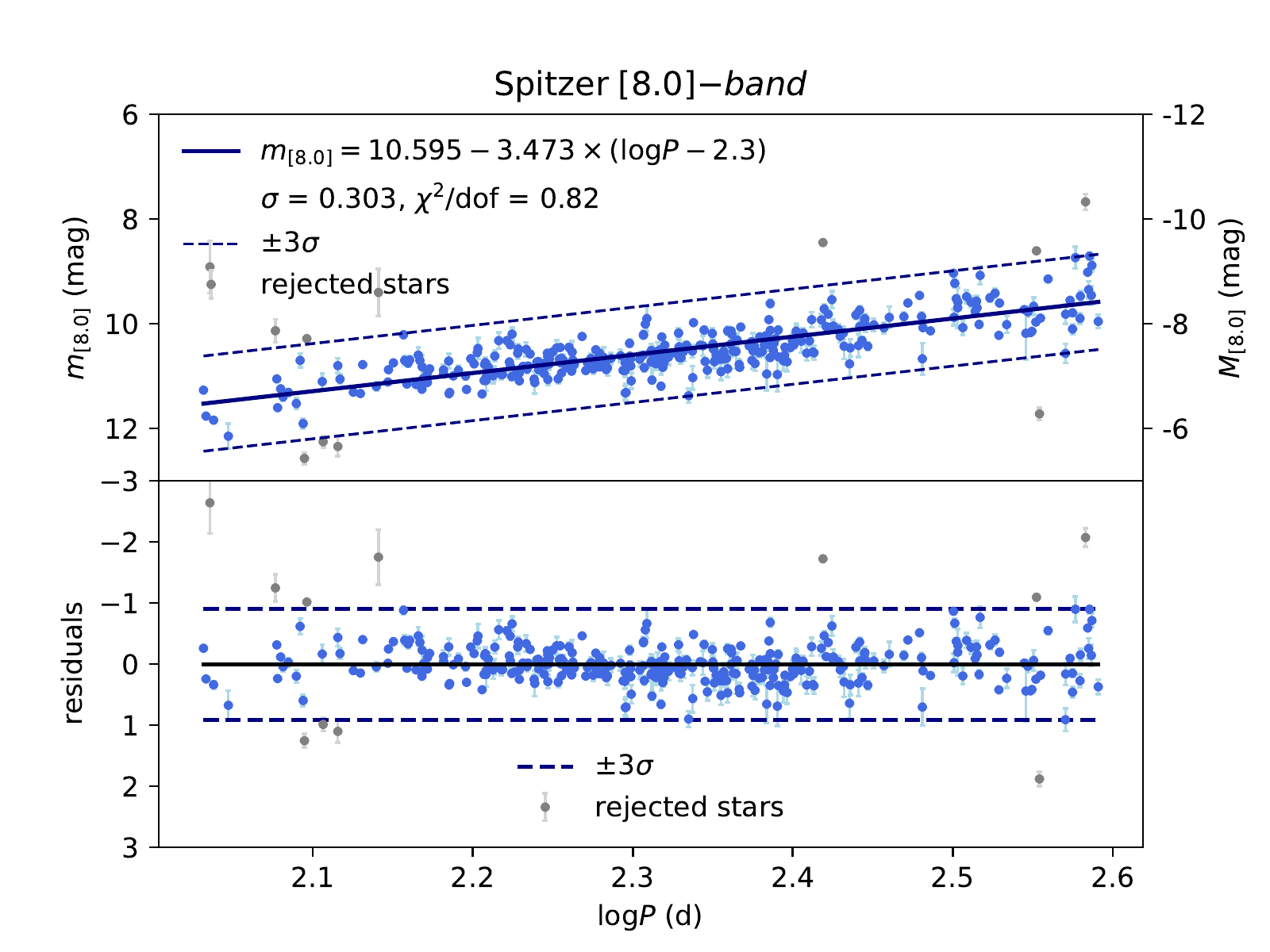}{0.5\textwidth}{(d)}
          }
\caption{Same as in Figure \ref{fig:pl_wise_o_rich_linear}, but for the Spitzer bands: (a) $[3.6] \; \mu m$, (b) $[4.5] \; \mu m$, (c) $[5.8] \; \mu m$, (d) $[8.0] \; \mu m$.}
\label{fig:pl_spitzer_o_rich_linear}
\end{figure*}

\begin{figure*}
\gridline{\fig{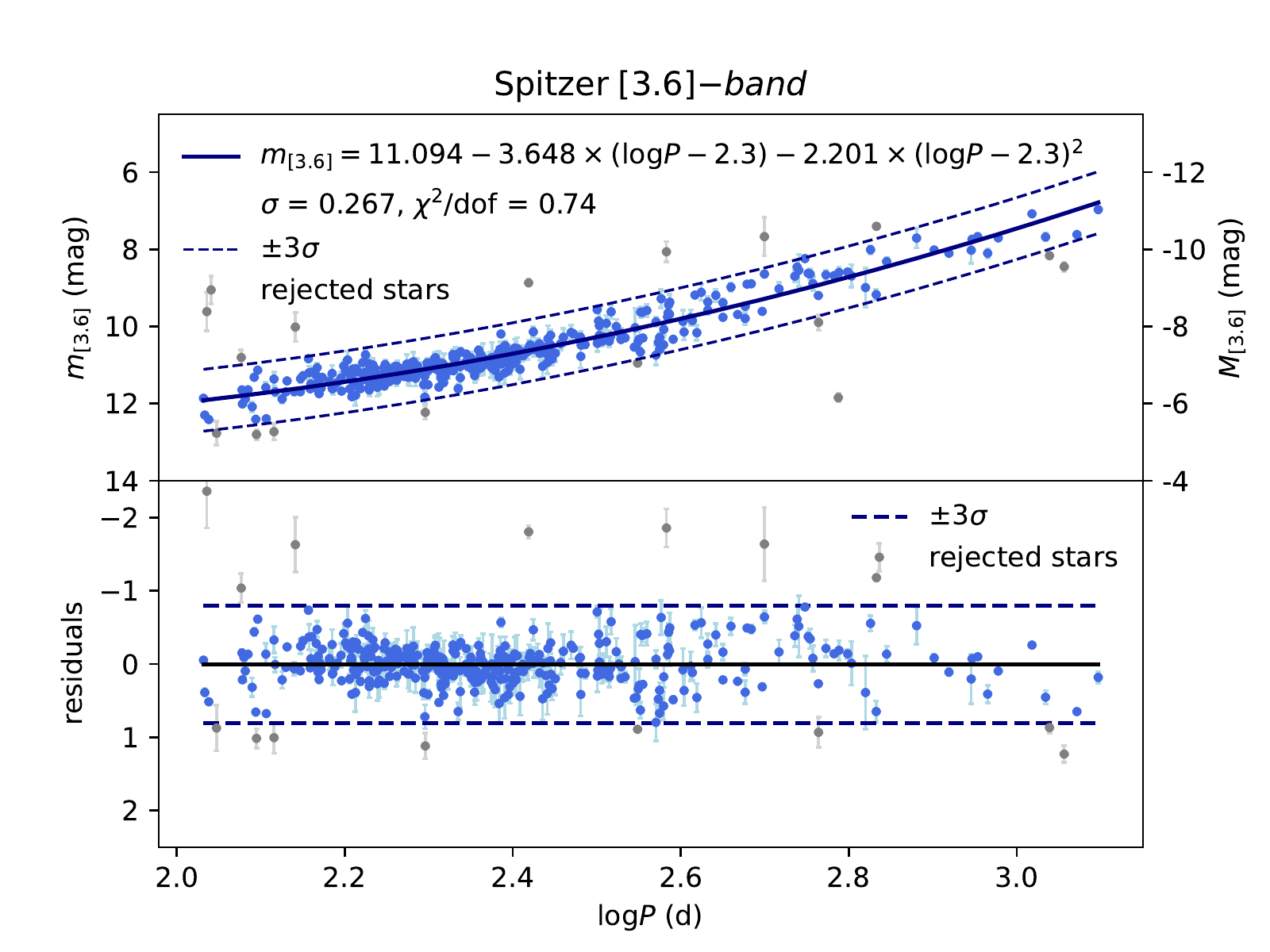}{0.5\textwidth}{(a)}
          \fig{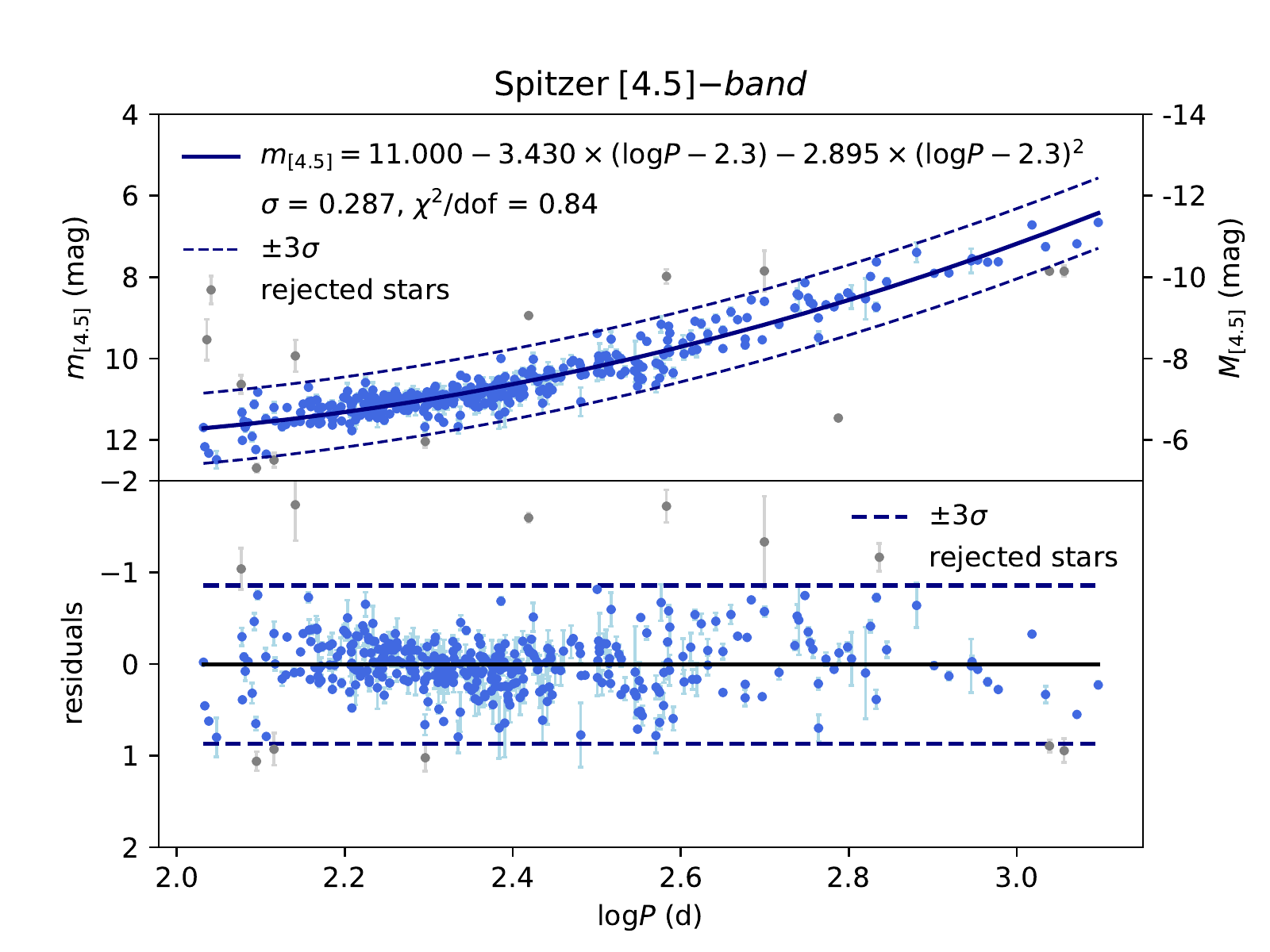}{0.5\textwidth}{(b)}
          }
\gridline{\fig{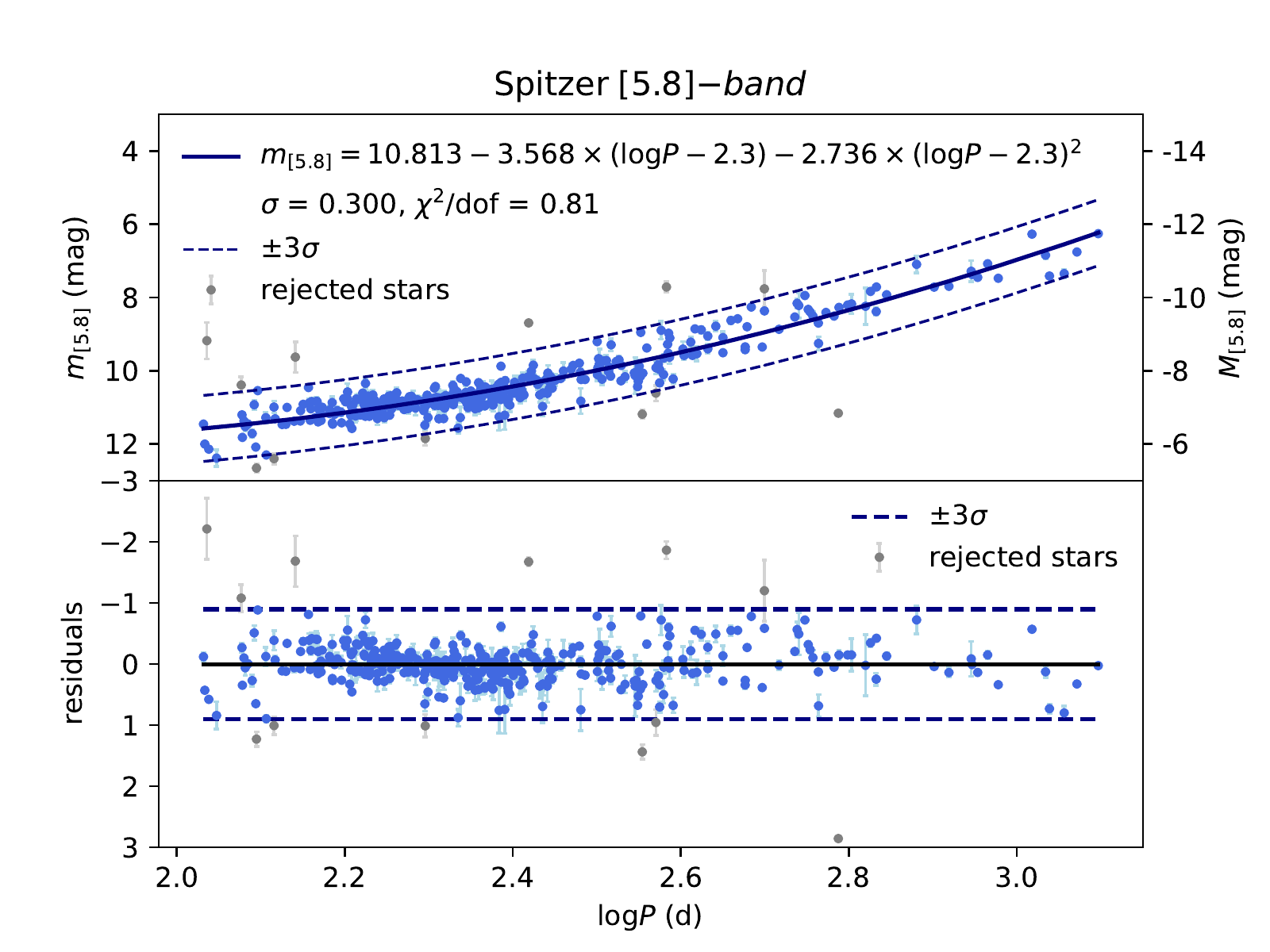}{0.5\textwidth}{(c)}
          \fig{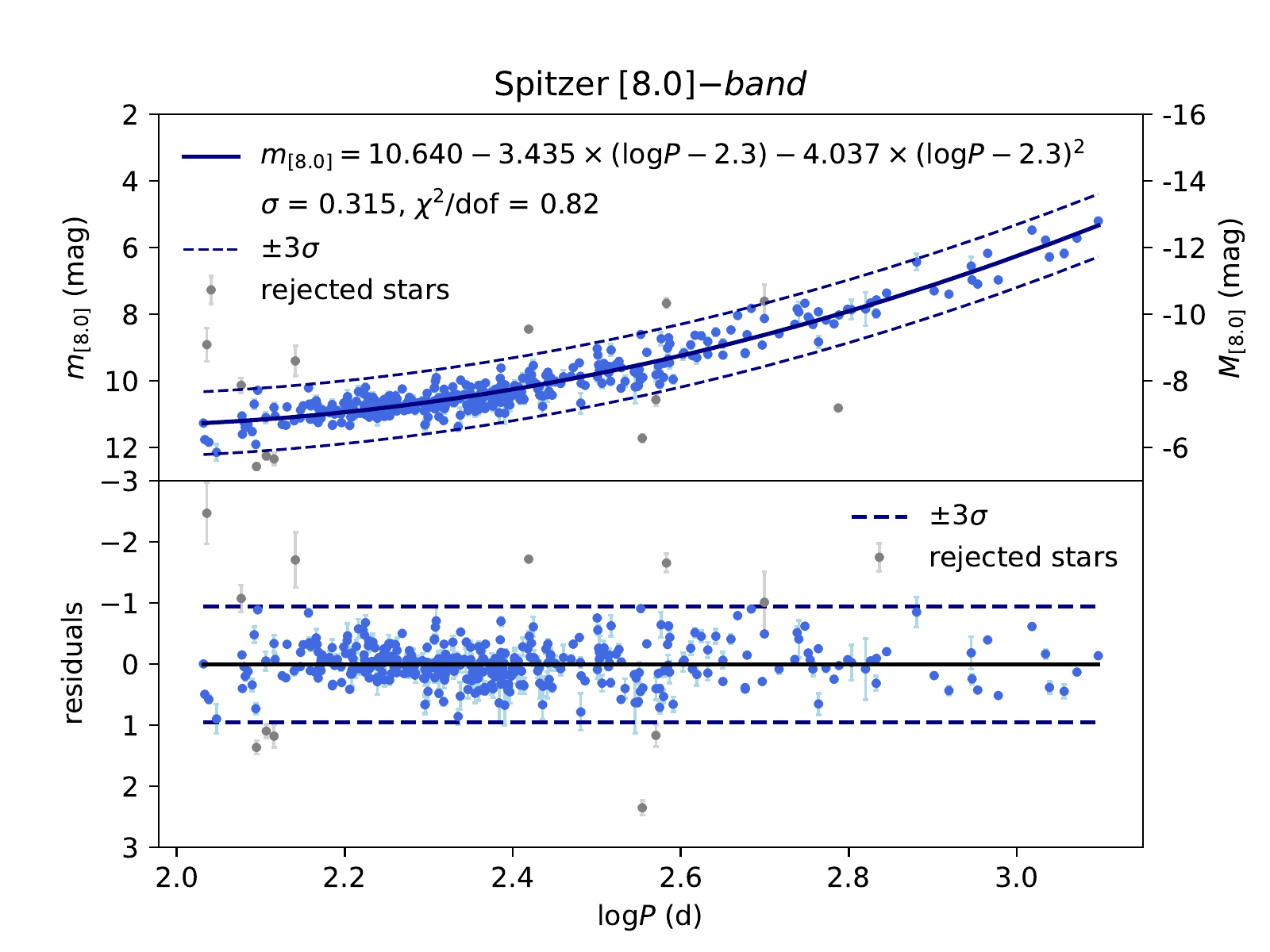}{0.5\textwidth}{(d)}
          }
\caption{Same as in Figure \ref{fig:pl_wise_o_rich}, but for the Spitzer bands: (a) $[3.6] \; \mu m$, (b) $[4.5] \; \mu m$, (c) $[5.8] \; \mu m$, (d) $[8.0] \; \mu m$.}
\label{fig:pl_spitzer_o_rich}
\end{figure*}

The PLRs in this paper (Figures \ref{fig:pl_wise_c_rich}--\ref{fig:pl_spitzer_o_rich} and Table \ref{tab:PL_results}) are based on the mean magnitudes measured from the fitted and scaled to mid-IR OGLE templates, as the integrated third-order Fourier series (see Section \ref{subsec:template_and_mean} in this paper, and \citealt{2021arXiv210703397I} for more details). In principle, in some cases this method could fail, as the Mira brightness in addition to typical periodic variability may exhibit aperiodic variations of the mean magnitude and amplitude. We compared these PLRs (in each Spitzer and WISE band) with PLRs constructed using mean magnitudes measured in two different ways. The PLRs fitting method is the same and is described in Section \ref{subsec:models_determining}.

In contrast to more sophisticated methods (i.e., the integrated Fourier series fit), we also measured the mean magnitudes using two simple methods: (1) the regular flux-based mean of the scaled OGLE templates to the mid-IR data, and (2) the sum of maximum and minimum fluxes divided by two and converted to magnitude. The magnitudes have been corrected for the interstellar extinction as described in Section \ref{subsec:extinction}. We then constructed PLRs separately for the C-rich Miras (in the full range of pulsation periods) and for the O-rich Miras (with linear fit to periods $\log P \leq 2.6$).

Let us compare the differences between the PLR parameters for the mean magnitudes obtained with different methods to the corresponding uncertainties of our base models (presented in Table \ref{tab:PL_results}). In both cases (base---method (1) and base---method (2)), the smallest differences from our base models were obtained for the mean magnitudes calculated as a regular mean (method (1)). For the \mbox{C-rich} Miras, the differences in zero-points are at the level of $0.5\sigma$, while for the O-rich Miras these differences are less than $0.5\sigma$, for both Spitzer and WISE data. The slope of PLRs varies around $3\sigma$ for our base models of C-rich Miras, while this differences for the O-rich Miras are smaller than $0.5\sigma$. Dispersions are the smallest, almost in all cases, for the mean magnitudes obtained from the integration of the Fourier series and are used in PLRs presented in Table \ref{tab:PL_results} and in Figures \ref{fig:pl_wise_c_rich}--\ref{fig:pl_spitzer_o_rich}. Method (2) significantly underestimates the mean magnitudes and causes much worse results than the base model.

\section{Conclusions}

In this paper, we used the catalog of Mira-type variable stars in the LMC from the OGLE survey \citep{2009AcA....59..239S} to construct the most accurate mid-IR PLRs to date. We cross-matched the known LMC Miras with the WISE \citep{2010AJ....140.1868W} and Spitzer databases \citep{2004ApJS..154....1W} to obtain mid-IR light curves. We used two-decades-long OGLE light curves to create templates using the GPR model \citep{2016AJ....152..164H}, and we fitted these templates to the mid-IR data \citep{2021arXiv210703397I}. We measured the mean magnitudes in mid-IR WISE and Spitzer bands by fitting a third-order truncated Fourier series, and integrating them. The mean magnitudes were corrected for the interstellar extinction using the reddening map of the LMC \citep{2021ApJS..252...23S}, and extinction curves in the \mbox{mid-IR} bands \citep{2018ApJ...859..137C, 2019ApJ...877..116W}. We calibrated the Mira PLRs using the most accurate distance measurement to the LMC \citep{2019Natur.567..200P}.

The PLRs presented in Table \ref{tab:PL_results} can be used to measure distances to the Mira-type stars in the Milky Way and nearby galaxies. The distance to an individual Mira star can be measured with the accuracy at the level of $5\%$ and $12\%$ for the O-rich and C-rich Mira, respectively. 

In future work, we are going to measure the \mbox{mid-IR} distances to the tens of thousands of Mira-type stars in the Milky Way discovered in the OGLE data. It will allow us to study the three-dimensional structure of the Milky Way.

\acknowledgments{
We are deeply grateful to the anonymous referee for the detailed report. The referee raised several points that turned out to be of significant importance. The suggestions given by the referee let us to greatly improved this manuscript, but more importantly, these comments motivated us to explored the subject of Miras variability in-depth. As a result, the referee's remarks led to the creation of a second paper entitled "Multiwavelength Properties of Miras" \citep{2021arXiv210703397I}.

We thank Drs. Jan Skowron, Przemek Mr\'oz, Radek Poleski, and Mariusz Gromadzki, and Marcin Wrona for comments that helped to improve this
manuscript. This work has been supported by the National Science Centre, Poland, grant MAESTRO no. 2016/22/A/ST9/00009 to IS. PI is partially supported by the {\it Kartezjusz} programme no. POWR.03.02.00-00-I001/16-00 founded by the National Centre for Research and Development, Poland. SK acknowledges the financial support of the Polish National Science Center through grant no. 2018/31/B/ST9/00334.

This publication makes use of data products from the {\it Wide-field Infrared Survey Explorer} (WISE), which is a joint project of the University of California, Los Angeles, and the Jet Propulsion Laboratory/California Institute of Technology, funded by the {\it National Aeronautics and Space Administration} (NASA). This work is based in part on archival data obtained with the Spitzer Space Telescope, which was operated by the Jet Propulsion Laboratory, California Institute of Technology under a contract with NASA.}

\onecolumngrid

\twocolumngrid

\onecolumngrid

\twocolumngrid

\onecolumngrid

\twocolumngrid

\clearpage

\bibliography{paper}{}
\bibliographystyle{aasjournal}

\end{document}